\documentclass[10pt]{article}
\usepackage[a4paper]{geometry}
\usepackage{graphicx,enumerate}
\usepackage{amsmath,amssymb}
\usepackage{authblk,cite,subcaption}
\usepackage{xfrac,authblk}

\usepackage{fancyhdr}
\usepackage[sc]{mathpazo}
\usepackage[T1]{fontenc}
\usepackage{hyperref}

\newcommand{\rmi}{\mathrm{i}} 
\newcommand{\rmd}{\mathrm{d}} 
\newcommand{\rme}{\mathrm{e}}
\newcommand{\kz}{\mathbf{k}_0}

\newcommand{\Xa}{\mathrm{X}} 
\newcommand{\Ya}{\mathrm{Y}} 
\newcommand{\Ma}{\mathrm{M}} 
\newcommand{\Ga}{\Gamma} 
   
\title{Trapped Modes  and Steered Dirac Cones  in Platonic Crystals}

\author[1]{R.C. McPhedran}
\author[1]{A.B. Movchan\thanks{Author for correspondence: abm@liverpool.ac.uk}}
\author[1]{N.V. Movchan}
\author[1,2]{M. Brun}
\author[3]{M.J.A. Smith}

\affil[1]{\small Department of Mathematical Sciences, University of Liverpool, Liverpool, L69 3BX, U.K.}
\affil[2]{Dipartimento di Ingegneria Meccanica, Chimica e dei Materiali, Universit\'{a} di Cagliari, Piazza d'Armi, I-09123 Cagliari, Italy}
\affil[3]{Institut Fresnel (UMR 7249), Aix--Marseille Universit\'e, Marseille 13013, France}

\date{\today}

\begin{document}

\maketitle

\begin{abstract}
This paper discusses the properties of flexural waves obeying the biharmonic equation, propagating in a thin plate pinned at doubly-periodic sets of points. The emphases are on the properties of dispersion surfaces having the Dirac cone topology, and on the related topic of trapped modes in  plates with a finite set (cluster) of pinned points. The Dirac cone topologies we exhibit have at least two cones touching at
a point in the reciprocal lattice, augmented by another band passing through the point. We show that the Dirac cones can be steered along symmetry lines in the Brillouin zone by varying the aspect ratio of rectangular lattices of pins, and that, as the cones are moved, the involved band surfaces tilt. We link Dirac points with a parabolic profile in their neighbourhood, and the characteristic of this parabolic profile decides the direction of propagation of the trapped mode in finite clusters.
\end{abstract}

\begin{center}
\end{center}

\section{Introduction}

There has been an explosion of interest in recent years in the properties and applications of topological insulators. These are naturally occurring materials for which the  structures of adjacent bands, associated with electrons obeying the Schr\"{o}dinger equation, can have the particular property of coming together at a single point called the Dirac point\cite{hasan2010colloquium}. This band geometry endows  materials such as graphene with a range of
properties conducive to applications, such as sensitivity to applied electric or magnetic fields and the ability to conduct electrons in surface states free from dissipation or backscattering, even when containing substantial amounts of impurities. These properties pertaining to electrons obeying the  Schr\"{o}dinger equation or particles obeying the Dirac equation can also be encountered in systems governed by Maxwell's equations, and is known as  topological photonics, as discussed by Lu {\it et al.} \cite{lu2014topological}.  They have also been exhibited for surface acoustic waves \cite{torrent2012acoustic} and in phononic crystals \cite{liu2011dirac}.

We will  present here  a discussion of topological platonics, showing how  equivalent band geometries can be generated for flexural waves obeying the biharmonic equation. For a particular case of two-dimensional lattices of pinned points, it was shown by Smith {\it et al.} \cite{smith2014double} that the same topological insulator geometry and band structure could be encountered in the centre of the Brillouin zone. Here, we will exploit the mathematical simplicity of the
band structures associated with the biharmonic equation for  arrays of pinned points to show how these interesting topologies can be steered along symmetry directions within the Brillouin zone by
employing rectangular rather than square or triangular lattices. The ratio of the periods  of the  rectangular lattice is an extra parameter, which can be employed to steer the Dirac point along the symmetry direction, in a way similar to that used in the control of filtering properties of stacked platonic gratings by Haslinger {\it et al.} \cite{haslinger2014symmetry}. The steering may be viewed simply, in that as the angle of wave propagation is altered, the separation of grating layers (or Fabry-Perot plates) needs to be adjusted  to keep the phase associated with propagation from one layer to the next `on resonance'.

The discussion of the topology associated with bands coming together at a single point was initiated by P. A. M. Dirac, who recognised that this geometry arose because the electrons (or particles) near the point propagated like waves in free space, unimpeded by any interaction with the atoms of the lattice (or with the quantum vacuum). This phenomenon for the Dirac equation is expressed concisely by saying the bands near the Dirac point correspond to massless particles. A second way of describing this effect is to  use the language of cloaking, and say that there are directions of neutral propagation around the Dirac point, where the wave propagates as if the crystal lattice were cloaked.
This property of neutrality for platonic crystals  was pointed out by McPhedran {\it et al.} \cite{mcphedran2009platonic}, and was explained in a simple way in terms of the singular directions of the Green's functions for the biharmonic equation.

The discussion of topological insulators distinguishes between Dirac points, where there are two perfect cones coming together at a point, and Dirac-like points, where this ideal geometry is perturbed \cite{mei2012linear}. In the case of topological platonics, the governing equation is of fourth order, not second order, and we show that, as well as the two cones, there is another dispersion  surface passing through the point. This surface corresponds to a wave influenced by interaction with the lattice, i.e. it is associated with a solution with non-zero mass.

In Section 2 of this paper, we give the governing equation for  flexural waves of the Kirchhoff type, and describe the associated lattice sums and Green's functions arising for two-dimensional lattices. In Section 3,  we discuss the connection between the rectangular lattice and the triangular lattice. For the latter, we describe the first two Dirac points, and the density of state (DOS) functions for flexural waves near each point.
These density of states functions are useful in describing the interaction between sources and flexural waves in the lattice structure.
In Section 4, we explicitly solve the equations under which two plane wave dispersion surfaces intersect for the rectangular lattice, and the equations under which three dispersion surfaces intersect.
The latter case gives the possible locations of Dirac points analytically. Section 5 discusses band diagrams and band surfaces for particular ratios of the two periods of the rectangular lattice; it is shown that ratios which are given by the square roots of integers exhibit particularly interesting behaviour. The contrast between massless and massive (non-zero mass) bands is highlighted, with this distinction depending upon the position of the Bloch vector in the Brillouin zone. Section 6 contains a discussion of flexural waves in  finite systems (clusters) of pinned points. We show that unidirectionally localised  modes of vibration can occur, and establish that their propagation direction corresponds to the single characteristic of a homogenised parabolic equation. Supplementary material to the paper includes animations showing the evolutions of band surfaces with varying aspect ratios of the rectangular lattice.

\section{Governing equation for thin plates}
We begin with the biharmonic model for flexural wave propagation in thin plates, which is given  by
\begin{equation}
(\Delta^2 - k^4)w(x,y) = 0.
\end{equation}
Here we have assumed a time dependence of $\exp (-\rmi \omega t)$ where $w$ denotes the out-of-plane displacement of the plate,  $\Delta$ denotes the two-dimensional Laplacian, $\omega$ is the angular frequency and $k$ is the nondimensionalised wave number which satisfies the dispersion relation $k^2 = \omega \sqrt{\rho_m h/D}$. For this model (see, for example Graff \cite{graff75}), $D=E h^3/[12 (1-\nu^2)]$ denotes the flexural rigidity of the plate, 
$E$ is Young's modulus, $\nu$ is Poisson's ratio, $h$ the thickness, and $\rho_m$ the mass density per unit area of the plate.

The zero-displacement condition for a pinned array of points is given by
\begin{equation}
\label{eq:vandisp}
w(\mathbf{R}_p) = 0,
\end{equation}
where $\mathbf{R}_p$ denotes the real-space  lattice vector corresponding to the  location of the pins.  Here we consider a rectangular lattice with periods $d_x$, $d_y$ along the $x$ and $y$ axes respectively, which gives rise to the complementary lattice vector pairs
\begin{subequations}
\begin{align}
\label{eq:RPRECT}
\mathbf{R}_p &=  \left[ \begin{array}{c} d_x p_x \\ 0 \end{array}\right] +  \left[ \begin{array}{c} 0 \\  d_y p_y \end{array}\right], \\
\mathbf{K}_h&= \frac{2\pi}{A} \left[ \begin{array}{c} h_x  d_y \\ 0 \end{array}\right] + \frac{2\pi}{A} \left[ \begin{array}{c} 0 \\ h_y d_x \end{array}\right], 
\end{align}
\end{subequations}
where $p=(p_x, p_y)$, $h=(h_x,h_y)$ denote  multi-indices with integer components,  $A=d_x d_y$ and $\mathbf{K}_h$ is the reciprocal lattice vector. For the pinned array problem we consider the plate displacement which is proportional to the quasiperiodic Green's function for the array. This Green's function is defined in terms of the rectangular lattice sums $\mathcal{R}_l$ which  can be expressed as 
\begin{subequations}
\begin{equation}
{\cal R}_l^Y=\sum_{p\neq (0,0)}Y_l (k R_p) \rme^{\rmi l \Theta_p}\rme^{\rmi{\mathbf k}_0\cdot {\mathbf R}_p},
\label{gl1}
\end{equation}
\begin{equation}
{\cal R}_l^K=\sum_{p\neq (0,0)}K_l (k R_p) \rme^{\rmi  l \Theta_p}\rme^{\rmi{\mathbf k}_0\cdot {\mathbf R}_p},
\label{gl2}
\end{equation}
\end{subequations}
where in the polar form $(r,\theta)$ we denote $\mathbf{R}_p=(R_p,\Theta_p)$.  The lattice sum $\mathcal{R}_l^Y$ is only conditionally convergent, and a suitable method of acceleration is required for effective numerical evaluation. Subsequently, we use the three-fold integrated expressions  \cite{movchan2007bloch}
\begin{subequations}
\begin{multline}
J_{l+3}(k\zeta) {\cal R}_l^Y=-\delta_{l,0}\left(Y_3(k\zeta)+\frac{1}{\pi}\sum_{n=1}^3 \frac{(3-n)!}{(n-1)!}\left( \frac{2}{k \zeta}\right)^{5-2 n}\right) \\
-\frac{4 \rmi^l}{A}\sum_h\left(\frac{k}{Q_h}\right)^3\frac{J_{l+3}(Q_h\zeta)
\rme^{\rmi l \Phi_h}}{(Q_h^2-k^2)},
\label{gl3}
\end{multline}
and
\begin{equation}
I_{l+3}(k\zeta) {\cal R}_l^K=\delta_{l,0}\left(K_3(k\zeta)-\frac{8}{(k\zeta)^3}+\frac{1}{k\zeta}-\frac{k\zeta}{8}\right)
+\frac{2\pi  \rmi^l}{A}\sum_h\left(\frac{k}{Q_h}\right)^3\frac{J_{l+3}(Q_h\zeta)
\rme^{\rmi l \Phi_h}}{(Q_h^2+k^2)}.
\label{gl4}
\end{equation}
\end{subequations}
  Here,  $\boldsymbol{\zeta}$ is an arbitrary vector lying inside the Wigner-Seitz cell (WSC) with length $\zeta <\min (d_x,d_y)$. We define ${\mathbf Q}_h={\mathbf k}_0+{\mathbf K}_h=(Q_h,\Phi_h)$ in polar form, and $A=d_x d_y$ for the rectangular array. In terms of these lattice sums, the quasiperiodic Green's function  is given by
\begin{multline}
G({\mathbf r};k,{\mathbf k}_0)=-\frac{1}{8 k^2} (Y_0(k r)+\frac{2}{\pi}K_0(k r)+\sum_{m=-\infty}^\infty \mathcal{R}_m^Y(k,{\mathbf k}_0)J_m(k r)e^{-\rmi m \theta} \\
+\frac{2}{\pi}\sum_{m=-\infty}^\infty \mathcal{R}_m^K(k,{\mathbf k}_0)I_m(k r)e^{-\rmi m \theta}),
\label{gl5}
\end{multline}
where ${\mathbf r}=(r,\theta)$ in polar form. From this representation we can recover (Movchan {\it et al.} \cite{movchan2007bloch}) the dispersion equation  via
 \begin{equation}
G({\mathbf 0};k,{\mathbf k}_0)=-\frac{1}{8 k^2}\left(\mathcal{R}_0^Y(k,{\mathbf k}_0)+\frac{2}{\pi} \mathcal{R}_0^K(k,{\mathbf k}_0)\right)= 0 .
\label{disprel2}
\end{equation}
 This elegant dispersion equation is a consequence of the assumption of zero radius for the pins constituting the periodic array. If we were to consider finite circular inclusions, the dispersion equation would then be evaluated as the determinant of a block matrix \cite{movchan2007bloch,poulton2010convergence}.

 As described in \cite{mcphedran2009platonic}, the flexural wave bands of platonic crystals may be "sandwiched" between  light lines in  reciprocal space which arise from singularities in the lattice sum \eqref{gl3}
 (when $Q_h=k$).
 In the case of rectangular arrays they take the particular form
 \begin{equation}
 (k_{0x}+2 \pi h_x/d_x)^2+ (k_{0y}+2 \pi h_y/d_y)^2=k^2,
 \label{tl1}
 \end{equation}
 for integers $h_x$, $h_y$. The sandwiching property means that points where light lines intersect constrain the dispersion curves of flexural waves in reciprocal space.

\section{Equivalences between lattice geometries: triangular and rectangular lattices}
In the following sections we explore the rectangular pinned lattice for various aspect ratios $\rho = d_y/d_x$. The Dirac cones that we observe include those for the square  and triangular (hexagonal) lattices. The fact that we see the Dirac cones for the triangular lattice follows from the correspondence between a subset of the points  of the triangular lattice and those of the rectangular lattice for $\rho = \sqrt{3}$. That is,  if we take a triangular lattice,  and form a sublattice by taking every even value of $p_y$ (see \eqref{triRPexpr} below), the result is equivalent to a rectangular lattice with aspect ratio $\rho = \sqrt{3}$. 
This equivalence is straightforward to observe if we numerically evaluate \eqref{eq:RPRECT} and the array vectors for the triangular lattice which are given by
 \begin{subequations}
\begin{align}
\mathbf{R}_{Tp}  &= p_x\left[ \begin{array}{c} d \\ 0 \end{array}\right] + p_y\left[ \begin{array}{c}d/2 \\ \sqrt{3} d /2 \end{array}\right].
\label{triRPexpr}
\end{align}
Consequently, the modes for the  triangular lattice may be viewed as the modes of two interlaced rectangular lattices. Given this equivalence between the rectangular and triangular arrays, we begin with an investigation of the Dirac points of a triangular lattice. Note that the triangular lattice is distinct from the  honeycomb lattice, the latter not being a Bravais lattice   and being generated by superposing two triangular arrays, i.e.  $R_p^{HC} = R_p^+ + R_p^-$ where
\begin{equation}
\mathbf{R}_p^\pm = d \left( m \left[ \begin{array}{c} \sqrt{3}/2 \\ 1/2\end{array} \right]  + n \left[ \begin{array}{c} \sqrt{3}/2 \\ -1/2\end{array} \right]      \right)  \pm \frac{d}{2\sqrt{3}} \left[ \begin{array}{c} 1  \\ 0\end{array} \right].
\end{equation}
 \end{subequations}
  Using the subscript $T$ to denote the triangular lattice, the dispersion equation for a triangular lattice of pinned points is
 \begin{equation}
G_T({\mathbf 0};k,{\mathbf k}_0)=-\frac{1}{8 k^2}\left({\cal R}_{T0}^Y(k,{\mathbf k}_0)+\frac{2}{\pi} {\cal R}_{T0}^K(k,{\mathbf k}_0)\right)=\frac{1}{A_{T}}\sum_h\frac{1}{Q_{Th}^4-k^4}=0,
\label{tr1}
\end{equation}
with $A_{T}$ being the area of the Wigner-Seitz cell of the triangular lattice.
Here the quasiperiodic reciprocal lattice vectors for the triangular lattice are given in Cartesian form  by
\begin{equation}
Q_{Th}=(k_{0x}+\frac{2\pi}{d_x}h_{T1},k_{0y}+(2 h_{T2}-h_{T1})\frac{2\pi}{\sqrt{3} d_x}).
\label{tr2}
\end{equation}
We can also consider the analogous expression corresponding to two shifted and superposed rectangular arrays 
\begin{subequations}
\begin{equation}
\frac{1}{A_{T}}\sum_h\frac{1}{Q_{h}^4-k^4} +\exp (i {\mathbf k}_0\cdot {\mathbf c}_{T}) \frac{1}{A_{T}}\sum_h\frac{e^{-i  {\mathbf c}_{T}\cdot{ \mathbf Q}_h} }{Q_{h}^4-k^4}=0.
\label{tr3}
\end{equation}
where ${\mathbf c}_{T}=d_x(1/2,\sqrt{3}/2)$. In the second term, the exponent simplifies to $-i  {\mathbf c}_{T}\cdot{ \mathbf K}_h$, or $-i \pi(h_x+h_y)$, so that (\ref{tr3}) becomes
\begin{equation}
\frac{1}{A_{T}}\sum_h\frac{1}{Q_{h}^4-k^4} \left(1+e^{-\pi i(h_x+h_y)}\right)=0.
\label{tr4}
\end{equation}
\end{subequations}
The sum in (\ref{tr4}) is thus restricted to integers $h_x$ and $h_y$ such that $h_x+h_y$ is even. Given that 
\begin{equation}
Q_{h}=\left( k_{0x}+\frac{2\pi h_{x} }{d_x},k_{0y}+\frac{2\pi h_y}{\sqrt{3} d_x} \right),
\label{tr5}
\end{equation}
the solutions to (\ref{tr4}) and (\ref{tr1}) are consistent, provided that  
\begin{equation}
\eta = G(\mathbf{c}_{T};k,{\mathbf k}_0)/G({\mathbf 0};k,{\mathbf k}_0)=\exp (i {\mathbf k}_0\cdot {\mathbf c}_{T}).
\end{equation}

 \begin{figure}[t]
\centering
\includegraphics[width=0.5\linewidth]{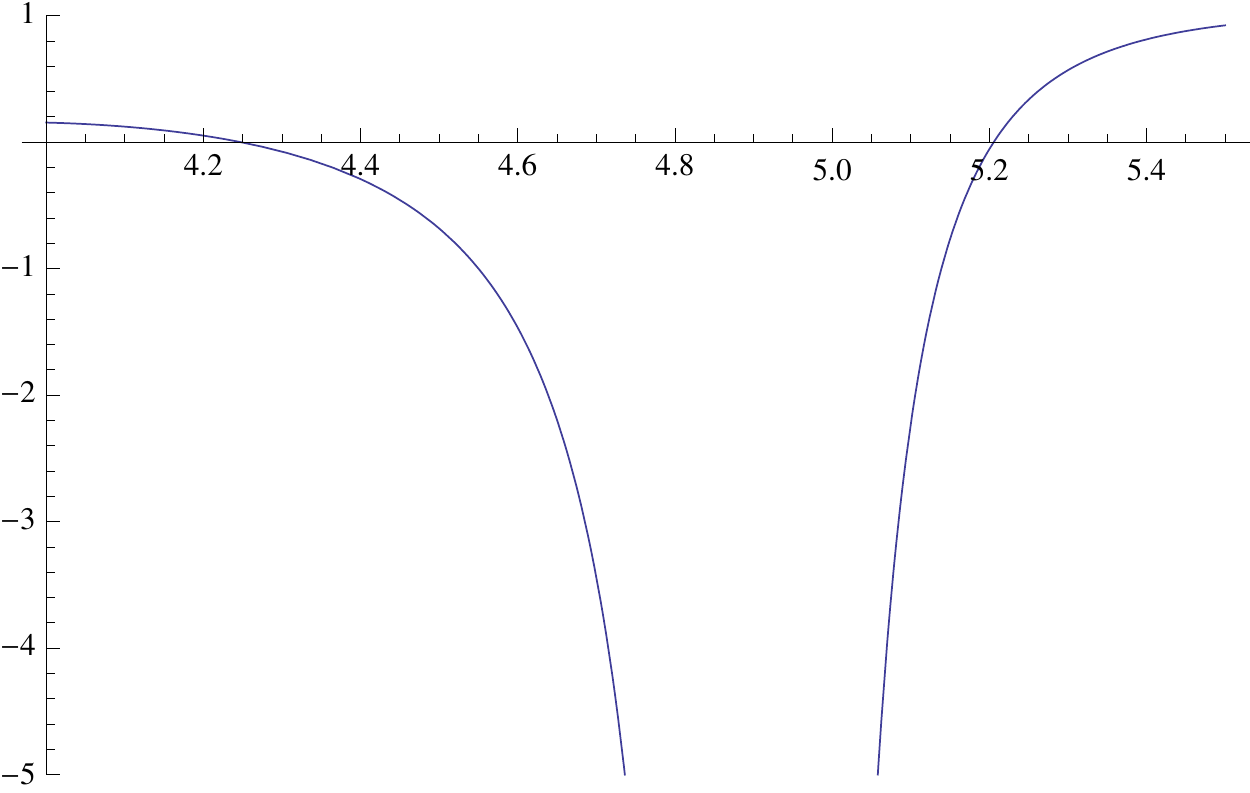}
\caption{The function $G({\mathbf 0};k,{\mathbf k}_0)$ as a function of $k$ for a triangular lattice \eqref{triRPexpr} where $d=1$,  $k_{0x}=k_{0y}=0$. Note the two zeros near $k=4.2$ and $k=5.2$. } \label{fig0}
\end{figure}

In figure \ref{fig0} we show the form of the dispersion line coming from equation \eqref{disprel2} as a function of $k$. In this case, there are two zeros  near $k=4.24743$ and $k=5.20563$. In the former case, the Bloch condition for $\eta$ is satisfied, whereas in the second it has a sign minus that required by the Bloch condition. Thus, only the former zero gives a solution  of the dispersion equation consistent with the Bloch condition for the triangular lattice.

 \subsection{The First Dirac Point}
\begin{figure}[t]
\centering
\includegraphics[width=.5\linewidth]{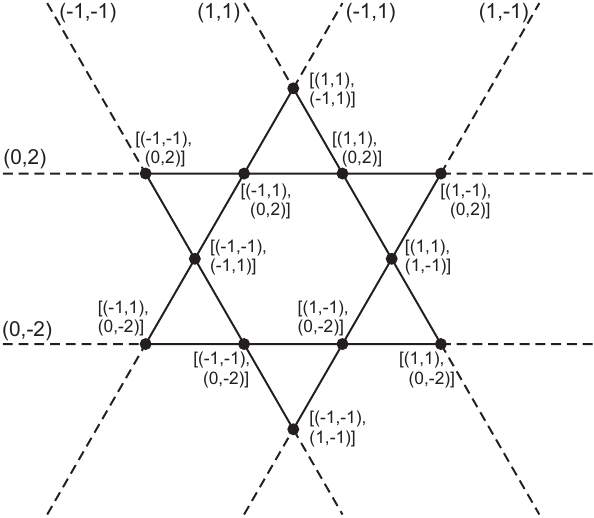}
\caption{Light lines labelled by the integers $(h_x,h_y)$ are shown in the vicinity  and above the first Dirac point at  $k=k_{D1}=(2 \pi / d_x) \sqrt{4/3}$. The dots at the intersection points define an outer and an inner Dirac cone,
which decrease in radius towards zero as $k$ tends to its value at the Dirac point.} \label{fig1}
\end{figure}

\begin{figure}[t]
\centering
\includegraphics[width=.5\linewidth]{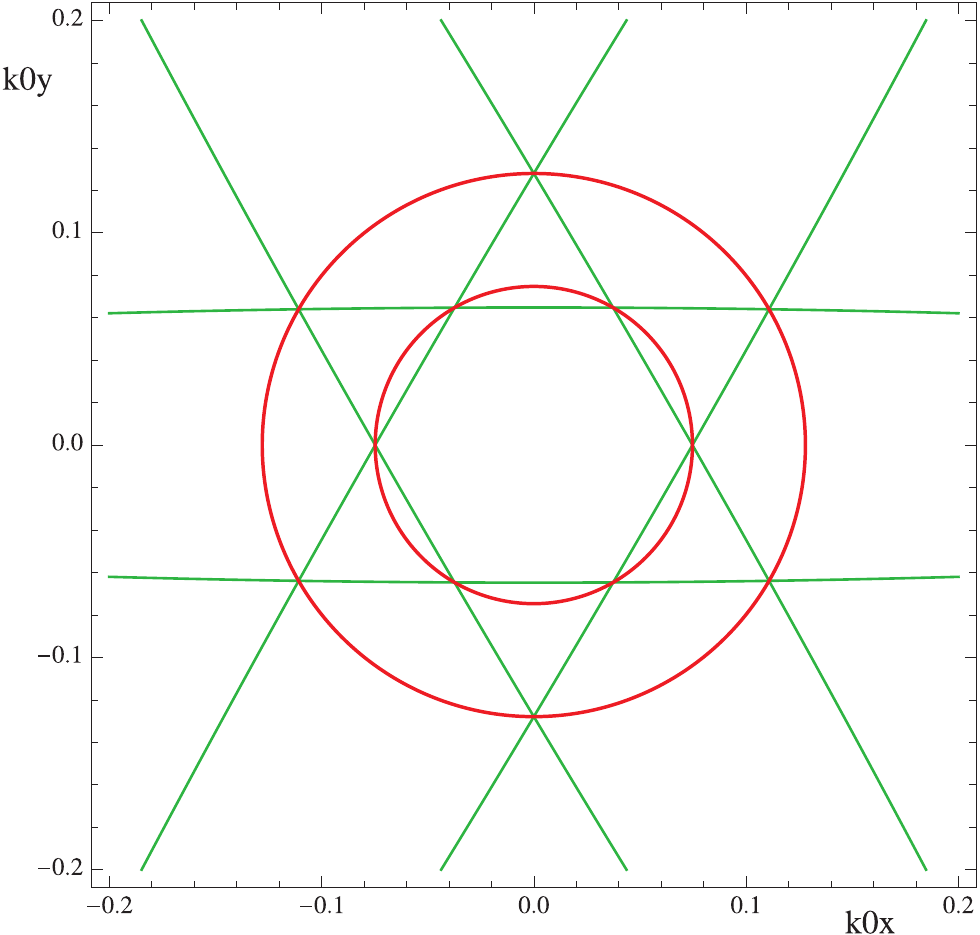}
\caption{Light lines are shown in the vicinity of the first
Dirac point at  $k=k_{D1}=(2 \pi / d_x) \sqrt{4/3}$. The red curves are circles of radii $2(k-k_{D1})$, $2(k-k_{D1})/\sqrt{3}$, which coincide to numerical
accuracy with the isofrequency contours for the triangular lattice for $k d_x=7.32$.\label{fig2}} 
\end{figure}

 Interesting Dirac cone topologies occurring near the origin in reciprocal space $(k_{0x},k_{0y})=(0,0)$
 were investigated by Smith {\it et al.} \cite{smith2014double}. For their case, the first Dirac point has $k=k_{D1}=(2 \pi / d_x) \sqrt{4/3}$. Figure \ref{fig1} shows the intersection points of the relevant light lines in the neighbourhood of the Dirac point.
 It can be seen that the light lines given by equation (\ref{tl1}) are well approximated by straight lines in the vicinity of the Dirac point.  The intersection point of the light lines defined by the indices $(1,1)$ and $(-1,1)$ is given by
\begin{subequations}
 \begin{equation}
 (k_{0x},k_{0y})=\left(0,~\frac{1}{3 d_x} \left(-2\sqrt{3}\pi+3\sqrt{k^2 d_x^2-4\pi^2}\right) \right),
 \label{tl2}
 \end{equation}
 which  when expanded about $k_{D1}$  becomes
 \begin{equation}
  (k_{0x},k_{0y})=\left(0,~ 2(k-k_{D1})-\frac{3 d_x\sqrt{3}}{4\pi} (k-k_{D1})^2 +O(k-k_{D1})^3 \right).
 \label{tl3}
 \end{equation}
\end{subequations}
 This gives the value $2$ for the radius of the outer Dirac cone (see figure \ref{fig2}).  The intersection point of the light lines $(1,1)$ and $(0,2)$ is
\begin{subequations}
 \begin{equation}
 (k_{0x},k_{0y})=\left(~\frac{1}{6 d_x} \left(-6 \pi+\sqrt{3}\sqrt{3 k^2 d_x^2-4\pi^2}\right),~\frac{1}{2 d_x} \left(-2\sqrt{3}\pi+3\sqrt{k^2 d_x^2-4\pi^2}\right) \right),
 \label{tl4}
 \end{equation}
 which  when expanded about $k_{D1}$  gives
 \begin{equation}
  \sqrt{k_{0x}^2+k_{0y}^2}=\frac{ 2}{\sqrt{3}}(k-k_{D1})-\frac{ d_x }{12\pi} (k-k_{D1})^2 +O(k-k_{D1})^3 .
 \label{tl5}
 \end{equation}
\end{subequations}
 Thus  the radius of the inner Dirac cone is $2/\sqrt{3}$ (see figure \ref{fig2}). As can be seen from figure \ref{fig1}, the six light lines $(\pm 1,\pm 1)$ and $(0,\pm 2)$ are all involved in defining the inner and outer Dirac cones.

\subsection{The Second Dirac Point}
\begin{figure}[ht!]
\centering
\begin{subfigure}[c]{0.42\textwidth}
\includegraphics[width=\linewidth]{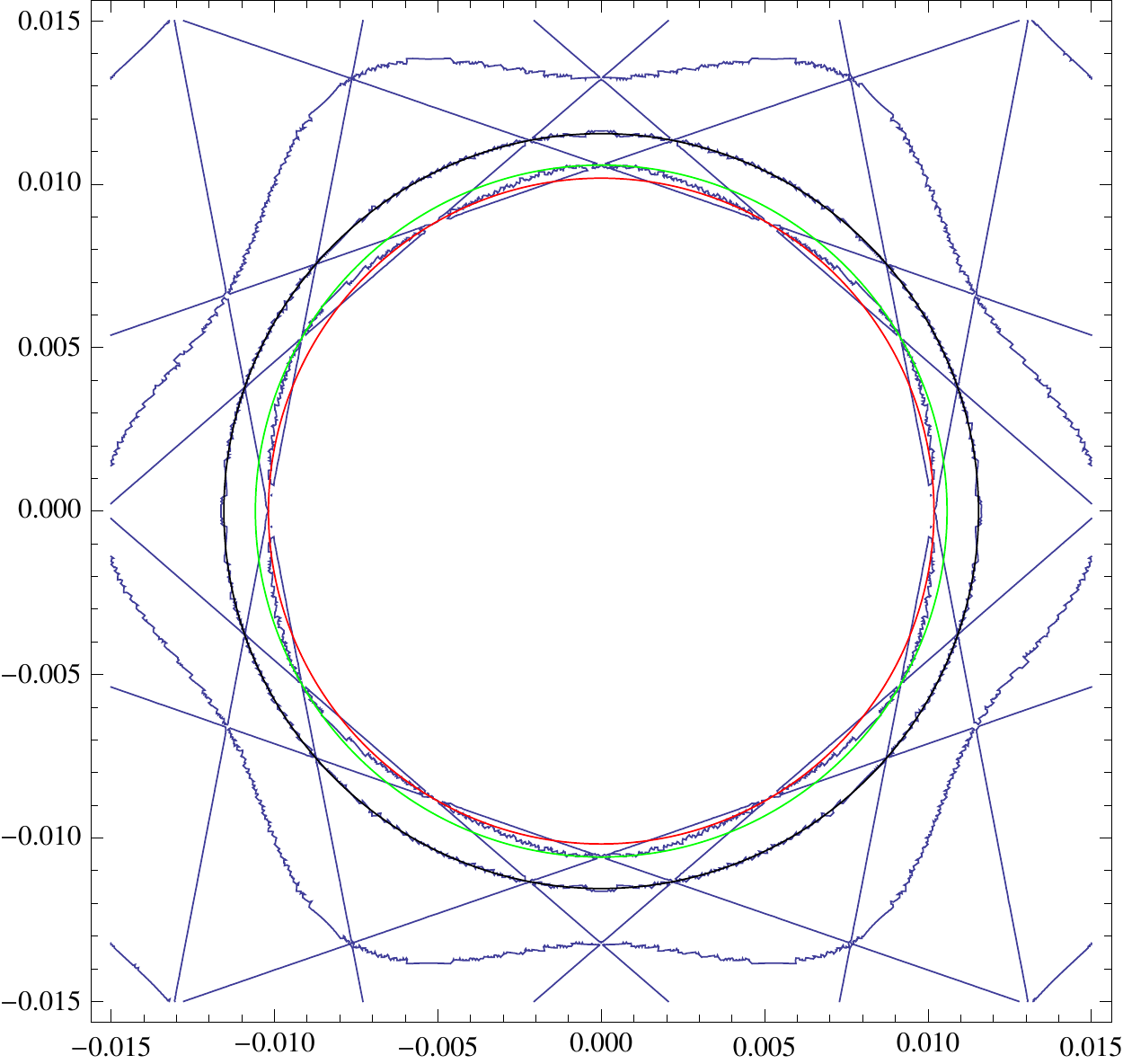}
\caption{\label{fig:bedtime1}}
\end{subfigure}
\qquad
\begin{subfigure}[c]{0.42\textwidth}
\includegraphics[width=\linewidth]{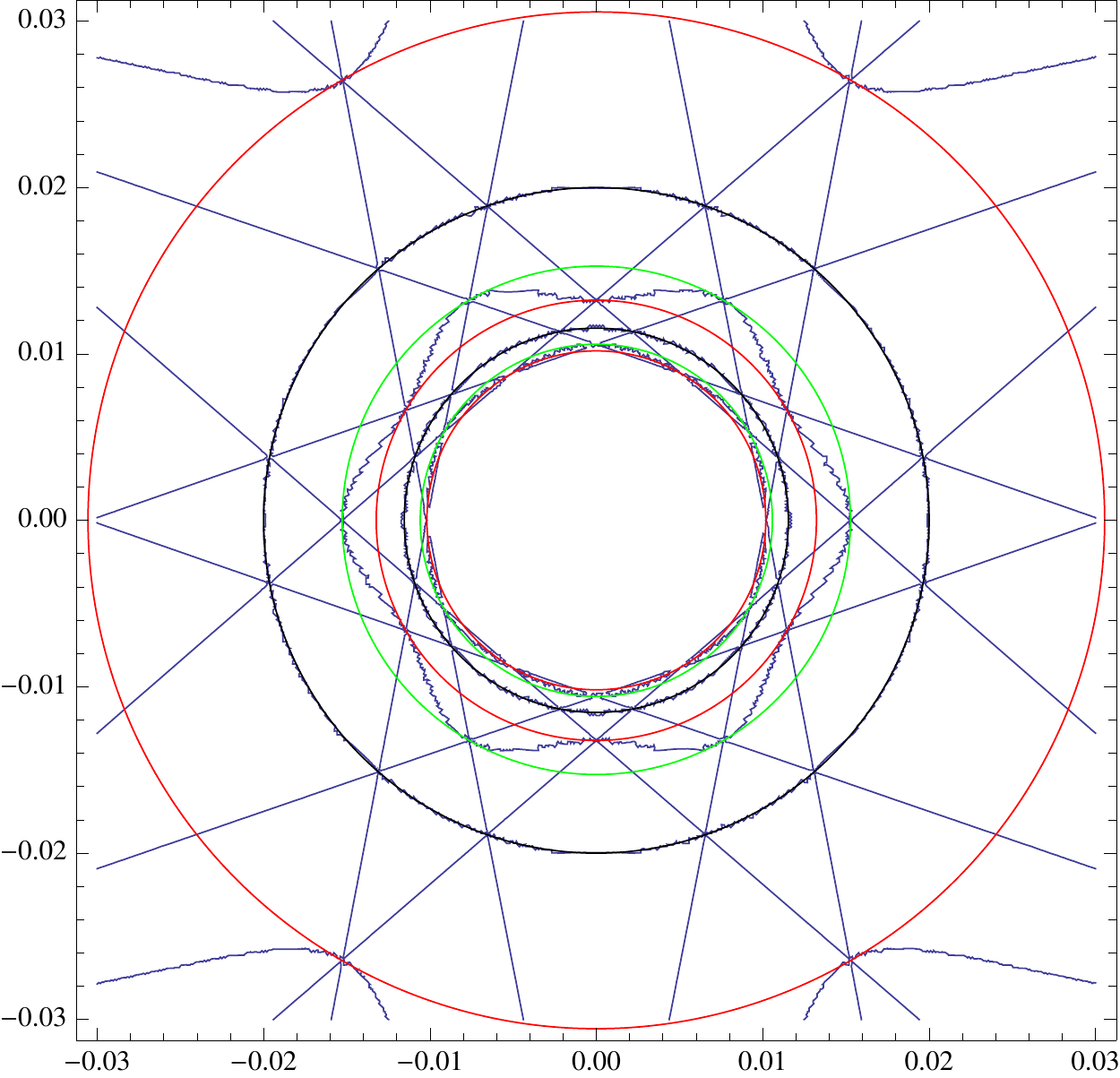}
\caption{\label{fig:bedtime2}}
\end{subfigure}
\begin{subfigure}[c]{0.42\textwidth}
\includegraphics[width=\linewidth]{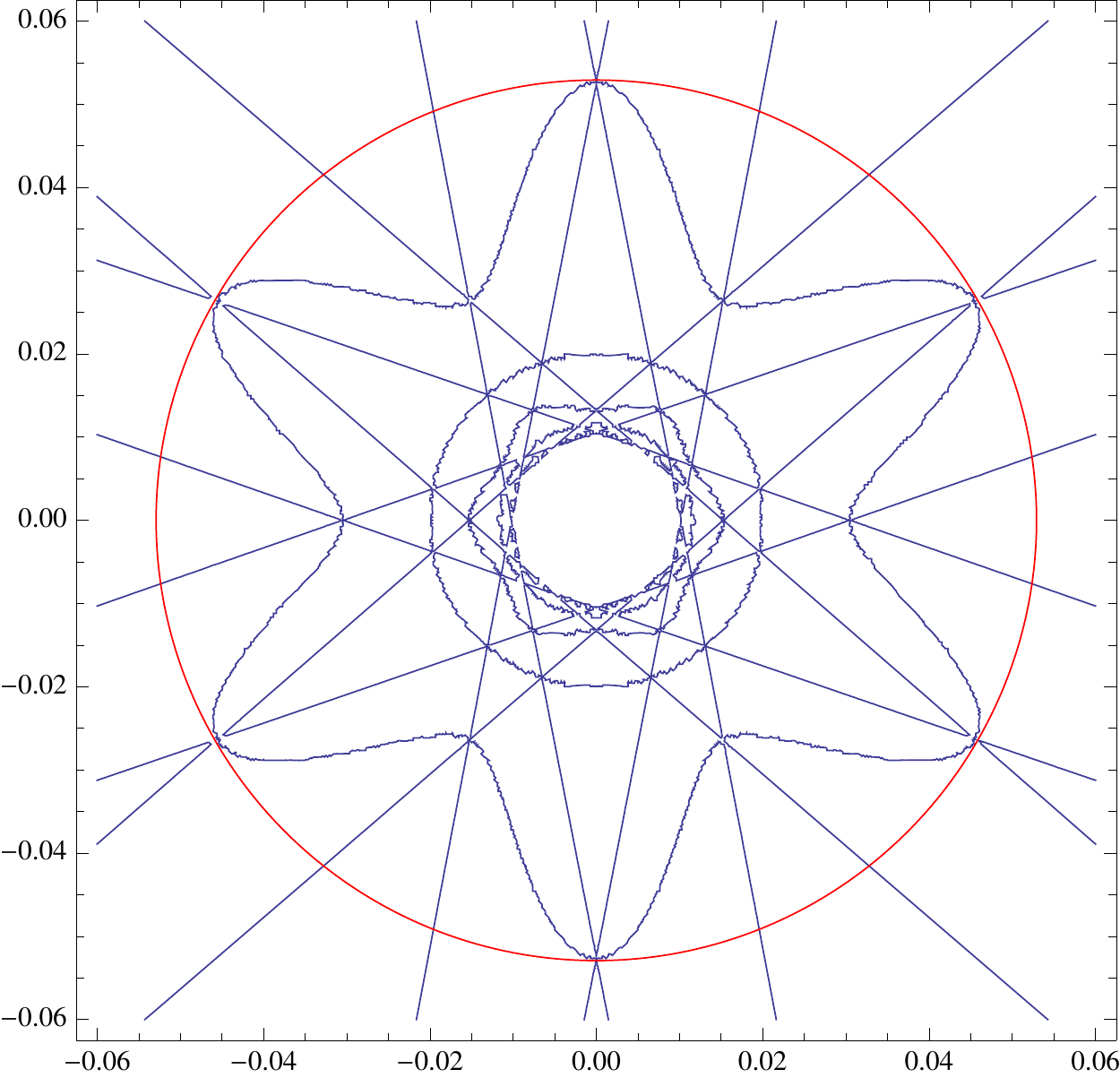}
\caption{\label{fig:bedtime3}}
\end{subfigure}
 
\caption{Light lines are shown in the vicinity of the Dirac point at  $k=k_{D2}=(2 \pi / d_x) \sqrt{28/3}$ with $k-k_{D2}=0.01$.
The isofrequency contours for the triangular lattice corresponding to $k d_x=19.20$ are given in blue.
 The coloured curves are circles of radii $0.01$ times those given in table \ref{table1} , which coincide to numerical accuracy with  The curves in other colours are circles of radius $0.01 \gamma_i$ as given in table 1 and bound the isofrequency contours. The   parts a b and c correspond to increasing ranges of $\Delta k$.\label{fig2iso}}
\end{figure}

The second Dirac point at $k_{0x} = k_{0y}=0$ has $k d_x=k_{D2} d_x=2\pi \sqrt{28/3}$. It is constructed by the intersection of twelve light lines rather than the six of the first Dirac point. These 
are defined by $(h_x,h_y)$ taking the values $(3,2)$, $(-3,-2)$, $(3,1)$, $(-3,-1)$, $(2,3)$, $(-2,-3)$, $(2,-1)$, $(-2,1)$, $(1,3)$, $(-1,-3)$, $(1,-2)$ and $(-1,2)$. Constructing the intersection points of each pair
of light lines, and choosing the root of smaller magnitude, we arrive at 48 pair intersections, from which their distances from the Dirac point are calculated. Also expanding the distance in terms of
$\Delta k=k-k_{D2} $ in the form 
\begin{equation}
\sqrt{k_{0x}^2+k_{0y}^2}=\alpha \Delta k -\beta (\Delta k)^2 + O(\Delta k^3),
\label{tl6}
\end{equation} 
we arrive at eight sets of coefficients, given in table \ref{table1}. The first two of these give in symbolic and numeric form the connection between the radius in the space of $(k_{0x},k_{0,y})$ and $\Delta k$. The third gives the curvature of the cone structure, whose sign is always such as to make the cone curve towards smaller radius.
\begin{table} \centering
\begin{tabular}{|c|c|c|c|c|}\hline
$\alpha$ & $\simeq \alpha$  & $\beta$ & Description  & $\gamma_i$ \\ \hline
$\frac{2}{3}\sqrt{\frac{7}{3}}$ & 1.01835 &$\frac{1}{324\pi}$ & Inner & $14 / 27\pi$ \\ \hline
$\frac{2\sqrt{7}}{5}$ & 1.05830 & $\frac{3\sqrt{3}}{500\pi} $ & Outer & $14/25 \pi$ \\ \hline
$\frac{2}{\sqrt{3}}$ & 1.15470 & $\frac{1}{12\sqrt{7}\pi}$ & Cone& $2/3\pi$ \\ \hline
$\frac{\sqrt{7}}{2}$ & 1.32288 & $\frac{3\sqrt{3}}{64\pi}$ & Inner & $7/8\pi$\\ \hline
$\sqrt{\frac{7}{3}}$ & 1.52753 & $\frac{1}{6\pi}$ & Outer & $1/2\pi$ \\ \hline
2 & 2.00000 & $\frac{3}{4\pi}\sqrt{\frac{3}{7}}$ & Cone  & $2/\pi$ \\\hline
$2\sqrt{\frac{7}{3}}$ & 3.05505 & $\frac{25}{12\pi}$ & Inner & $14/9\pi$ \\ \hline
$2\sqrt{7}$ & 5.29150 & $\frac{27\sqrt{3}}{4\pi}$ & Outer  & $14/\pi$ \\ \hline
\end{tabular}
\caption{ The coefficients $\alpha$, giving the cone radius, and $\beta$, giving the curvature, for the eight elements of the structure around the second Dirac point at $(k_{0x},k_{0y})=(0,0)$. The fourth column denotes whether the cone is a bounding cone (inner or outer) for a noncircular contour, or an actual conical band surface (cone). The final column gives the coefficient of
$|k-k_{D2}|$ in the  contribution to the DOS from each cone.}
\label{table1}
\end{table}
As can be seen in figure \ref{fig2iso}, the structure surrounding the second Dirac point consists of three distorted circles, with two intervening circles. The former are characterised by inner and outer radii, and the latter by twelve intersection points all lying at the same radius. If we define a proportionate distortion as the difference between the outer and inner radii, divided by their mean, then the proportionate distortion increases by the ratio $2+\sqrt{3}$  between the first and second distorted circles, and also between the second and third.

\subsection{Density of states near the first Dirac point}
The conical band surfaces above and below the first Dirac point  are sandwiched by light lines along certain directions in reciprocal space. This fact allows us to obtain the group velocity for the band analytically, and therefore the density of states (DOS). The light lines for a triangular lattice are given by
\begin{equation}
k(\kz) = |\mathbf{Q}_h| = \sqrt{\left( k_{0x} + 2 m \pi\right)^2 + \left( k_{0y} + \frac{4 \pi}{\sqrt{3}} (n-m/2) \right)^2}.
\label{eq:dos1}
\end{equation}
In particular, the inner cone  below the first Dirac point is sandwiched along $\Gamma \mathrm{K}$, or along the path $\kz = (t/2,\sqrt{3} t/2)$ between the light lines given by $(m,n) = (-1,-1)$ and $(m,n) = (0,-1)$. Using this parametrisation and either of these index pairs, we can establish that the slope of the cone is given by
\begin{subequations}
\begin{equation}
\partial_t k = \frac{t - 2\pi}{\sqrt{t^2 - 4 \pi t + 16 \pi^2 /3}}, \quad \mbox{ and } \quad \partial_t k\big|_{t=0} = -\frac{\sqrt{3}}{2}.
\label{eq:dos2}
\end{equation}
From \cite{smith2014dos} it is known that $|\mathbf{v}_g| = 2 k |\partial_t k|$, and so at $t=0$, $|\mathbf{v}_g| = \sqrt{3} k$. The outer cone below the Dirac point is sandwiched between the light lines with indices $(-1,-1)$ and $(1,0)$ along the path $\Gamma \mathrm{M}$ or $\kz = (0,t)$, which admits
\begin{equation}
\partial_t k = \frac{t - 2\pi/\sqrt{3}}{\sqrt{ 4\pi^2 + (t - 2\pi/\sqrt{3})^2 }}, \quad \mbox{ and } \quad \partial_t k\big|_{t=0} = -\frac{1}{2},
\label{eq:dos3}
\end{equation}
\end{subequations}
with a corresponding group velocity at $t=0$, $|\mathbf{v}_g| =   k$. It has been shown previously that the leading coefficient for the radius of the outer circle near the first Dirac point is given by $r_1 = 2|k-k_{D1}|$, and that for the inner circle we have $r_2 = 2|k - k_{D1}|/\sqrt{3}$. From  \cite{smith2014dos},  the DOS can be expressed as
\begin{equation}
\mathcal{N}(k) = \frac{2k}{A_{WSC} A_{BZ}} \sum_{m} \int_{C_m} \frac{1}{|\mathbf{v}_g|} \rmd s,
\label{eq:dos4}
\end{equation}
and, since the product of the areas of the Wigner Seitz cell and the Brillouin zone is $4\pi^2$ in two dimensions,
\begin{equation}
\mathcal{N}_{below}(k) = \frac{k}{2\pi^2} \left[ \frac{2\pi r_1}{\sqrt{3} k} + \frac{2\pi r_2}{k} \right] = (\gamma_1 + \gamma_2)|k-k_{D1}| = \frac{4}{\sqrt{3}\pi}|k-k_{D1}|.
\label{eq:dos5}
\end{equation}
 where $\gamma_i$ represents the contribution to the DOS from the $i^\mathrm{th}$ band surface. These are given by $\gamma_1 =  \gamma_2 = 2 / \sqrt{3}\pi$.  The central flat band (trapped between the light lines with indices $(-1,0)$ and $(1,0)$ along $\Gamma \mathrm{K}$) has
\begin{equation}
\partial_t k = \frac{t}{\sqrt{  16 \pi^2/3 +t^2 }}, \quad \mbox{ with } \quad \partial_t k\big|_{t=0} = 0,
\label{eq:dos8}
\end{equation}
and  $|\mathbf{v}_g| = 0$ at $t=0$, however the DOS is not unbounded as the length of the isofrequency contour at $k=k_{D1}$ is zero. A careful examination of this limit shows that the central band contributes a Heaviside step function at $k=k_{D1}$.

\subsection{Density of states at the second Dirac point}
We now proceed to an investigation of the DOS near the second Dirac point of the triangular lattice at  $k_{D2}(\Gamma) = 2 \pi \sqrt{28/3}$, where we have eleven degenerate bands  collapsing to a single point. For this point, only some of these bands are perfect cones. By carefully examining the light line surfaces about this point, it can be shown that the noncircular contours are bound  by circles as we approach the Dirac point. This can be seen in figure \ref{fig2iso}, with  radii and curvatures listed in table \ref{table1}.

If we assume that the contribution to the DOS from each curved surface is equal to the contribution from their bounding cones, we can repeat the procedure used above for the first Dirac point, to obtain an analytical expression for the DOS as we approach the second Dirac point. For example, for the lowest band surface (which is nonconical) we have an inner radius given by a light line crossing in the direction $\kz = t(1/2,\sqrt{3}/2)$ involving the indices $(m,n) = (-1,-3)$ and $(-2,-3)$. At this point  the slope of either light line is given by
\begin{subequations}
\begin{equation}
\partial_t k|_{t=0} = -\frac{3 \sqrt{3/7}}{2},
\label{eq:nyll1}
\end{equation}
with a corresponding group velocity at $t=0$, $|\mathbf{v}_g| =  3 k \sqrt{3/7}$. This inner circle has a radius of $r_1 = 2/3\sqrt{7/3}|k-k_{D2}|$ and thus the leading-order contribution to the DOS is given by
\begin{equation}
\mathcal{N}_1 = \frac{k}{2\pi^2} \frac{2 \pi r_1}{ |\mathbf{v}_{g1}| } = \gamma_1 |k-k_{D2}|  = \frac{14}{27\pi}|k-k_{D2}|.
\label{eq:ny11a}
\end{equation}
The outer cone has a light line crossing in the direction $\kz = (t,0)$ involving the indices $(-1,-3)$ and $(1,-2)$, giving rise to the slope
\begin{equation}
\partial_t k|_{t=0} = -\frac{5}{2\sqrt 7},
\label{eq:nyll2}
\end{equation}
and at $t=0$, $|\mathbf{v}_g| =  5k/\sqrt{7}$, with a radius $r_2 = 2\sqrt{7}/5|k-k_{D2}|$ giving rise to  the leading-order  contribution
\begin{equation}
\mathcal{N}_2 = \frac{k}{2\pi^2} \frac{2 \pi r_2}{ |\mathbf{v}_{g2}| } =  \gamma_2|k-k_{D2}| =\frac{14}{25\pi}|k-k_{D2}|.
\label{eq:ny12b}
\end{equation}
\end{subequations}

\section{Dispersion and Degenerate Light Lines}
We illustrate solutions of the dispersion equation $G({\mathbf 0};k,{\mathbf k}_0)=0$ for the rectangular lattice, and in particular, we will be interested in the conditions required to create degenerate light lines for the rectangular lattice. The light line equation for the mode $(n,m)$ is
\begin{equation}
\left(k_{0x}+\frac{2\pi n}{d_x}\right)^2+\left(k_{0y}+\frac{2\pi m}{d_y}\right)^2=k^2.
\label{degll1}
\end{equation}
We define rescaled, dimensionless quantities: $\kappa_x=k_{0x} d_x/(2 \pi)$, $\kappa_y=k_{0y} d_y/(2 \pi)$, with $0\leq \kappa_x,\kappa_y\leq 1/2$ corresponding to the first quadrant of the Brillouin zone.
 If $K=(k d_x)/(2 \pi)$ and $\rho = d_y/d_x$, then the light line equation (\ref{degll1}) becomes
\begin{equation}
\rho^2\left(n+\kappa_x\right)^2+\left(m+\kappa_y\right)^2=\rho^2 K^2.
\label{degll2}
\end{equation}
The condition for the light line for mode $(n,m)$ to coincide with that for mode $(n',m')$ is then
\begin{equation}
2  (m'-m) \kappa_y=2 \rho^2 (n-n') \kappa_x+\rho^2 (n^2-n'^2) +  (m^2 -m'^2).
\label{degll3}
\end{equation}
This is a straight line in the reciprocal space, which is only of interest if it lies within the Brillouin zone. The straight line is independent of the value of $K$ or $k$, but given the point $(\kappa_x,\kappa_y)$
on it, $K$ and $k$ can be determined from  equation (\ref{degll2}) or equation (\ref{degll1}). Note the special case $m=m'$, for which
\begin{subequations}
\begin{equation}
\kappa_x=-\frac{(n+n')}{2}.
\label{degll4}
\end{equation}
Allowed  solutions in this special case are:
\begin{equation}
\kappa_x=-\frac{1}{2},~ (n',m')=(1-n,m);~~\kappa_x=0,~ (n',m')=(-n,m);~~\kappa_x=\frac{1}{2},~ (n',m')=(-1-n,m).
\label{degll5}
\end{equation}
\end{subequations}
There is a second special case, $n=n'$, in which
\begin{subequations}
\begin{equation}
\kappa_y=-\frac{(m+m')}{2}.
\label{degll6}
\end{equation}
Allowed  solutions in this second special case are:
\begin{equation}
\kappa_y=-\frac{1}{2},~ (n',m')=(n,1-m);~~\kappa_y=0,~ (n',m')=(n,-m);~~\kappa_y=\frac{1}{2},~ (n',m')=(n,-1-m).
\label{degll7}
\end{equation}
\end{subequations}

\begin{figure}[ht!]
\centering
\includegraphics[width=7cm]{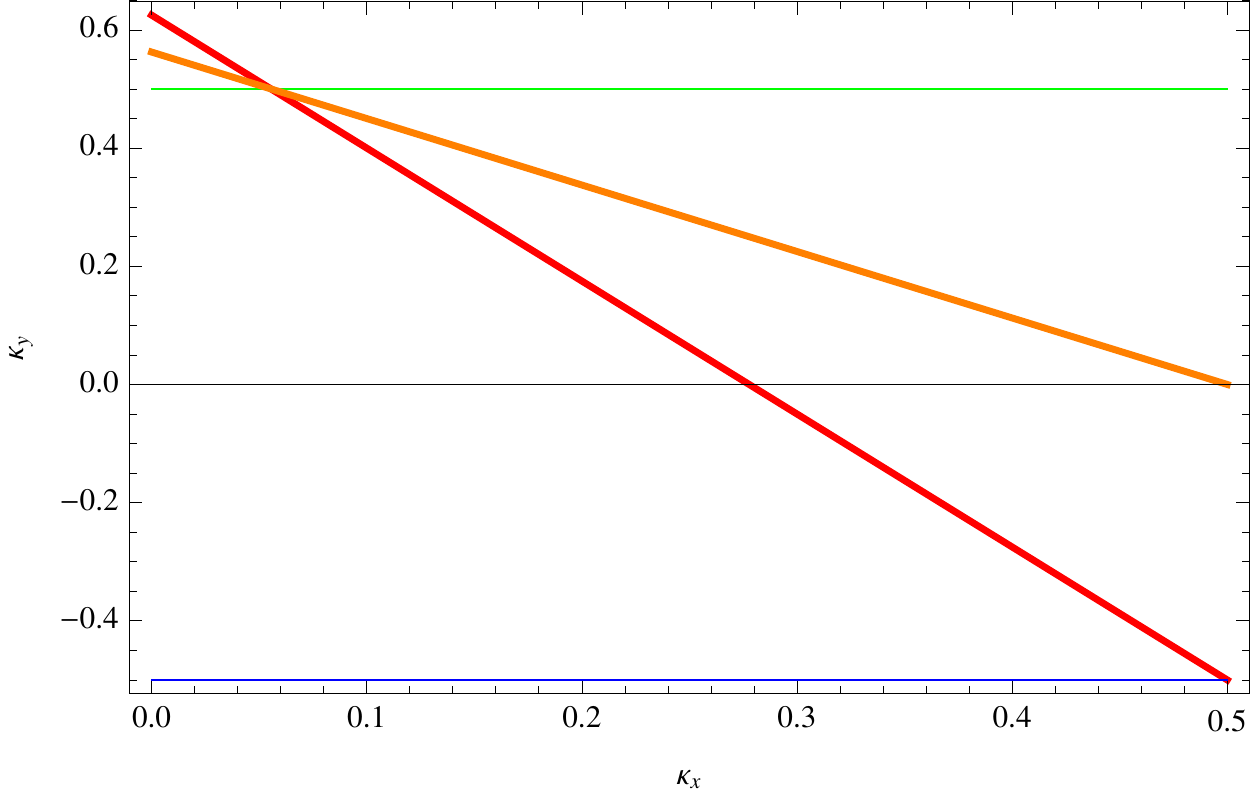}
\caption{Degenerate light line trajectories in the first quadrant of the Brillouin zone for $d_x=1$, $d_y=3/2$. The line colours correspond to the following integer quartets $(n,m,n',m')$:
red (-1, 0, 0, 1); green  (-1, 0, -1, -1); blue  (-1, 0, -1, 1);dashed brown (and obscured by the central line)  (-1, -1, -1, 1); orange  (-1, -1, 0, 1).\label{figdegll}} 
\end{figure}

\begin{figure}[t]
\centering
\begin{subfigure}[c]{0.45\textwidth}
\includegraphics[width=\linewidth]{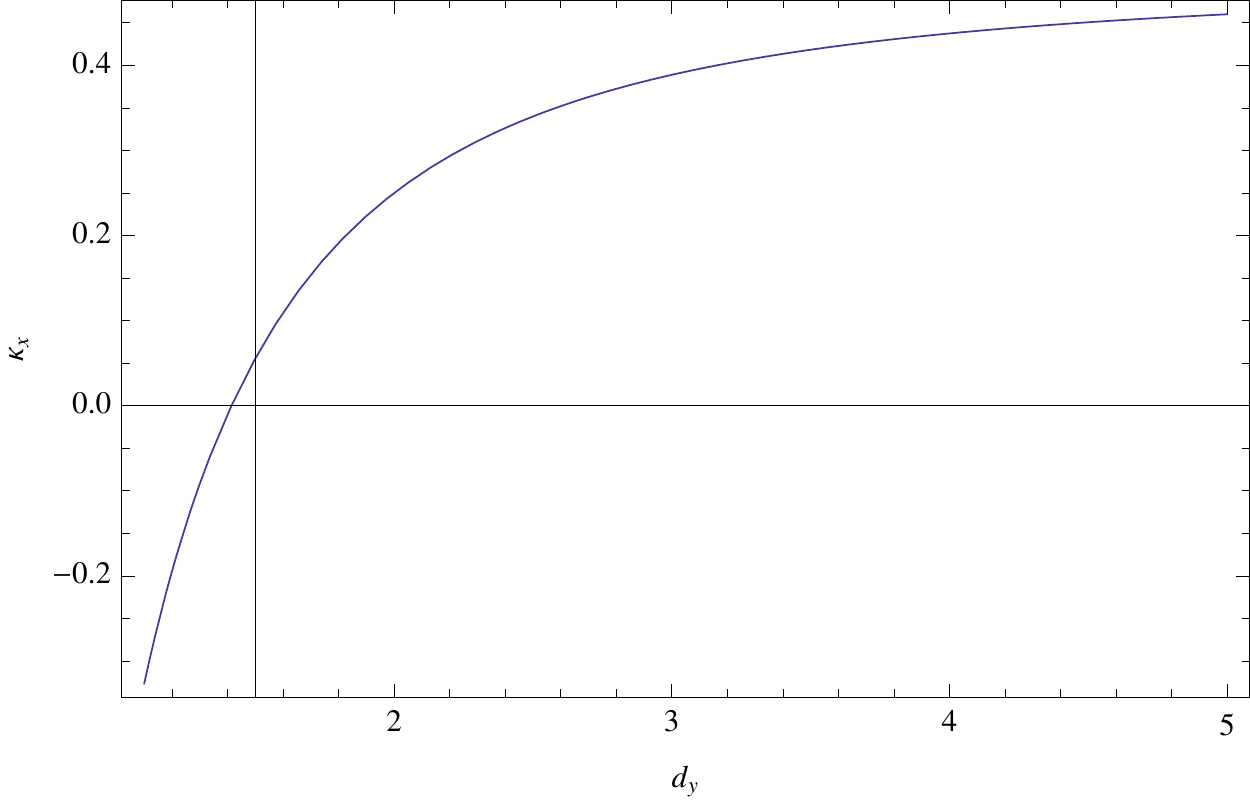}
\caption{\label{fig:bedtime1a}}
\end{subfigure}
\qquad
\begin{subfigure}[c]{0.15\textwidth}
\includegraphics[width=\linewidth]{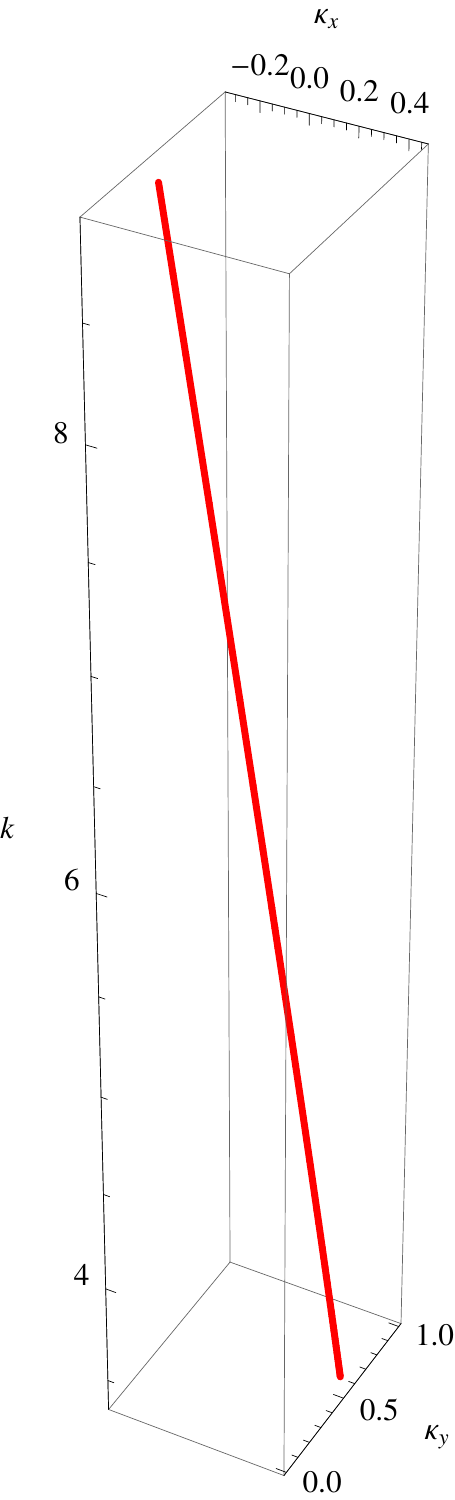}
\caption{\label{fig:bedtime2a}}
\end{subfigure}
\caption{\subref{fig:bedtime1a} $\kappa_x$ as a function of $d_y$ with $d_x=1$ for the triple intersection of light lines specified by the sextet $(n,m,n^\prime,m^\prime,n^{\prime \prime},m^{\prime \prime})=(-1,0,0,1,-1,-1)$.
\subref{fig:bedtime2a} a parametric plot giving $k$ as a function of  $\kappa_x$ and  $\kappa_y$, the parameter being $d_y$, which ranges from 1.1 to 5.\label{tripkapx}} 
\end{figure}

As an example of the use of (\ref{degll4}) we show in figure \ref{figdegll} degenerate light line trajectories corresponding to five integer quartets. Note the triple intersection at the top of the plot.
It corresponds to $\kappa_x=1/18$, $\kappa_y=1/2$ and $k=(5\sqrt{13}/9)\pi$. The numerical value of $k$ is approximately $2.00308 \pi$. We next find the intersection of the degenerate light lines corresponding to the quartets $(n,m,n',m')$ and $(n,m,n'',m'')$. These intersect at a point in the Brillouin zone for which
\begin{subequations}
\begin{equation}
\kappa_x = \frac{N_x}{D_x},
\label{degll8}
\end{equation}
with
\begin{multline}
N_x = m^\prime ( m^{\prime \prime 2} - m^2) +m^{\prime \prime  }(m^2 - m^{\prime   2} ) + m(m^{\prime   2} - m^{\prime \prime 2}) + \\
 \rho^2 \left[ n^{  \prime 2} ( m - m^{\prime \prime }) + n^{\prime \prime 2} (m^{\prime } - m) + n^2(m^{\prime \prime  } - m^{\prime  }) \right],
\end{multline}
and
\begin{equation}
D_x = 2 \rho^2 \left[ m(n^{\prime \prime} - n^\prime) + m^\prime (n - n^{\prime \prime}) + m^{\prime \prime} (n^\prime - n) \right].
\end{equation}
\end{subequations}
Also,
\begin{subequations}
\begin{equation}
\kappa_y = \frac{N_y}{D_y},
\label{degll9}
\end{equation}
with
\begin{equation}
N_y = n ( m^{\prime \prime 2} - m^{\prime 2}) +n^\prime (m^2 - m^{\prime \prime  2} ) + n^{\prime \prime} (m^{\prime   2} - m^{  2}) +  
 \rho^2 \left[ (n - n^\prime)(n - n^{\prime \prime})(n^\prime - n^{\prime \prime})  \right],
\end{equation}
and
\begin{equation}
D_y = D_x/\rho^2=2 \left[ m(n^{\prime \prime} - n^\prime) + m^\prime (n - n^{\prime \prime} ) + m^{\prime \prime} (n^\prime - n)   \right] .
\end{equation}
\end{subequations}
At such points at least three light lines (and possibly more) intersect, with the latter possibility easily verified given the knowledge of $(\kappa_x,\kappa_y,k)$.
Figure \ref{tripkapx} shows two plots generated using the expressions (\ref{degll8}), (\ref{degll9}). The first gives $\kappa_x$ as a function of $d_y$, with for the case in question $\kappa_y=1/2$ independent of $d_y$.
The value of $d_y$ for which $\kappa_x$ is zero is $\sqrt{2}$.  The second plot is a parametric curve along which the parameter $d_y$ changes from $1.1$ to $5$, with $\kappa_x$, $\kappa_y$ and $k$ all varying 
according to $d_y$.

\begin{figure}[t!]
\centering
\begin{subfigure}[c]{0.9\textwidth}
\includegraphics[width=\textwidth]{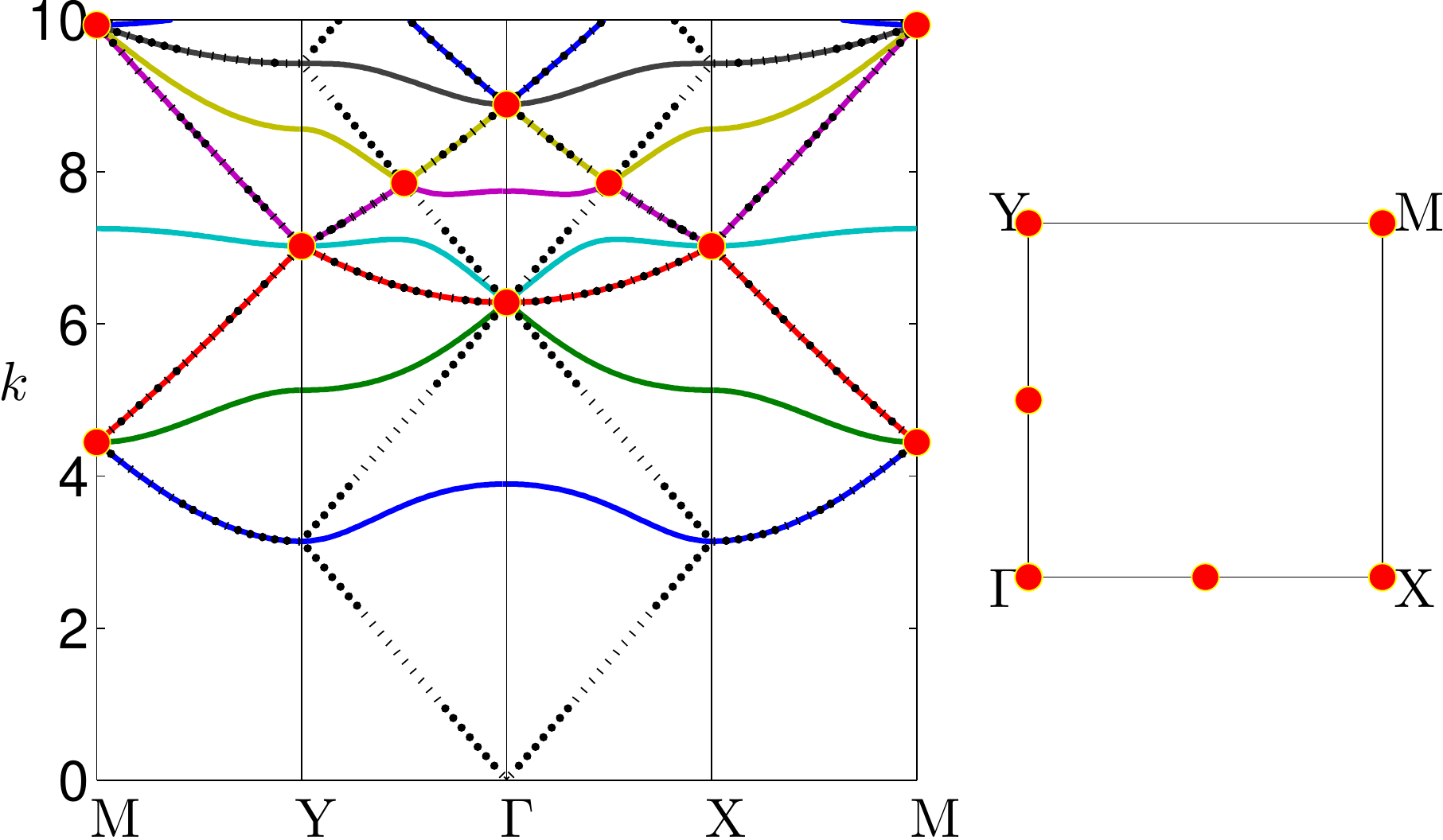}
\caption{\label{fig:bdiag1}}
\end{subfigure} \\
\begin{subfigure}[c]{0.9\textwidth}
\includegraphics[width=\textwidth]{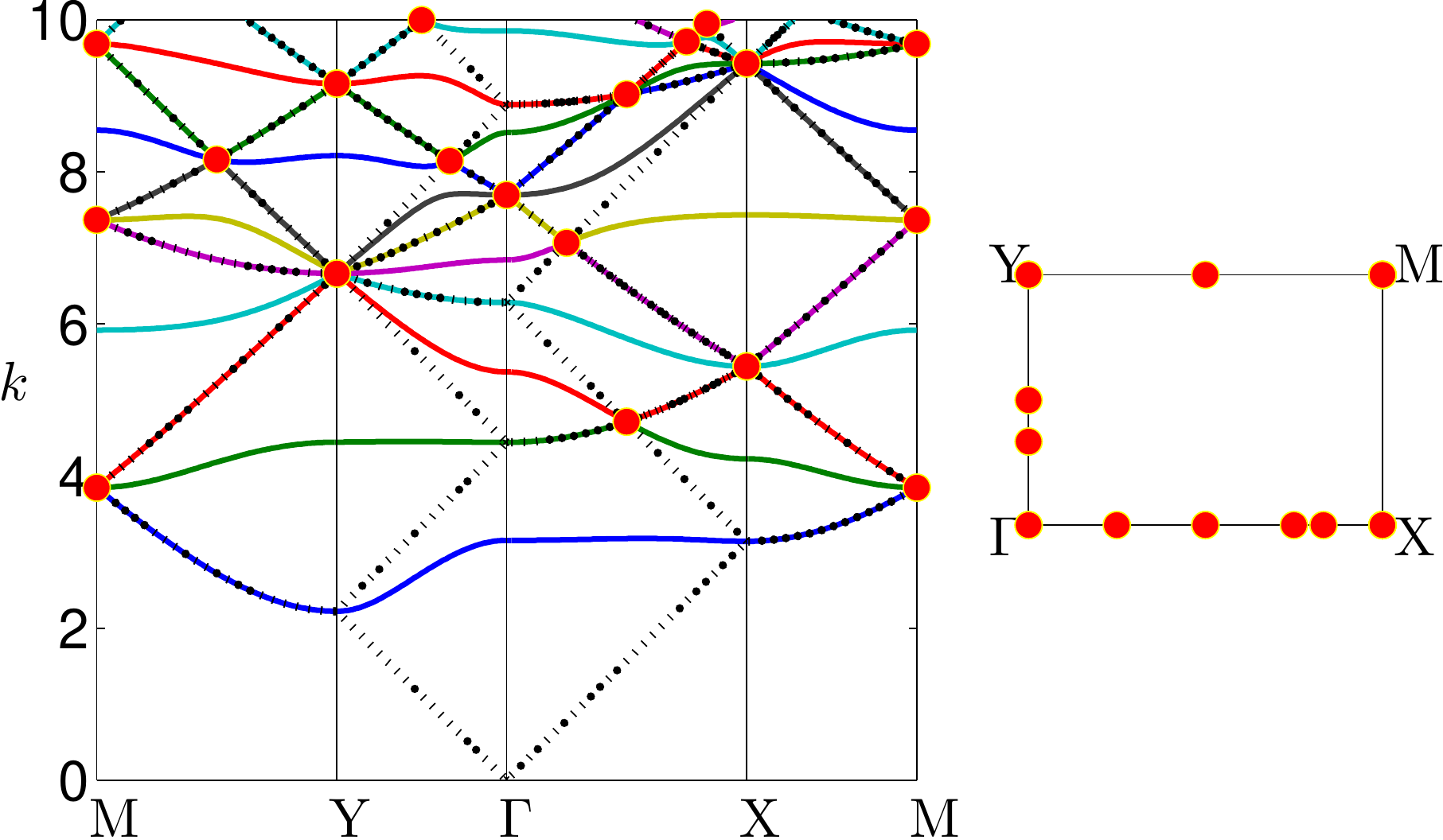}
\caption{\label{fig:bdiag2}}
\end{subfigure}
\caption{The band diagrams for \subref{fig:bdiag1} $\rho = 1$, and \subref{fig:bdiag2} $\rho = \sqrt{2}$, as we trace a path around the edge of the BZ for a rectangular array geometry. The points of light line triple intersections, given by \eqref{degll8} and \eqref{degll9}, are highlighted with red dots, and inset is the top-down projection of these dots to the edges of the zone.\label{bdiagb1}   } 
\end{figure}

\begin{figure}[t!]
\centering
\begin{subfigure}[c]{0.9\textwidth}
\includegraphics[width=\textwidth]{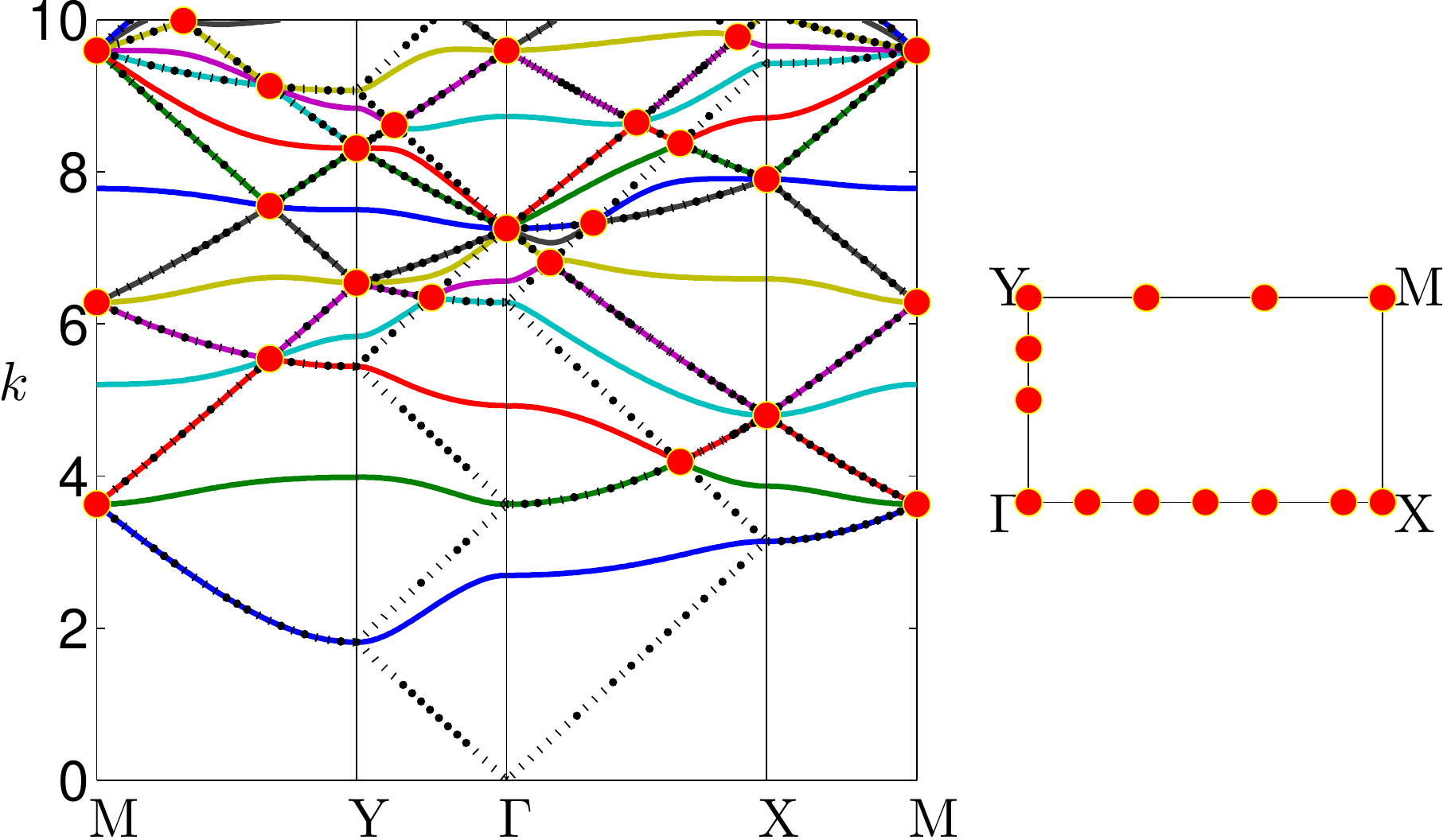}
\caption{\label{fig:bdiag3}}
\end{subfigure} \\
\begin{subfigure}[c]{0.9\textwidth}
\includegraphics[width=\textwidth]{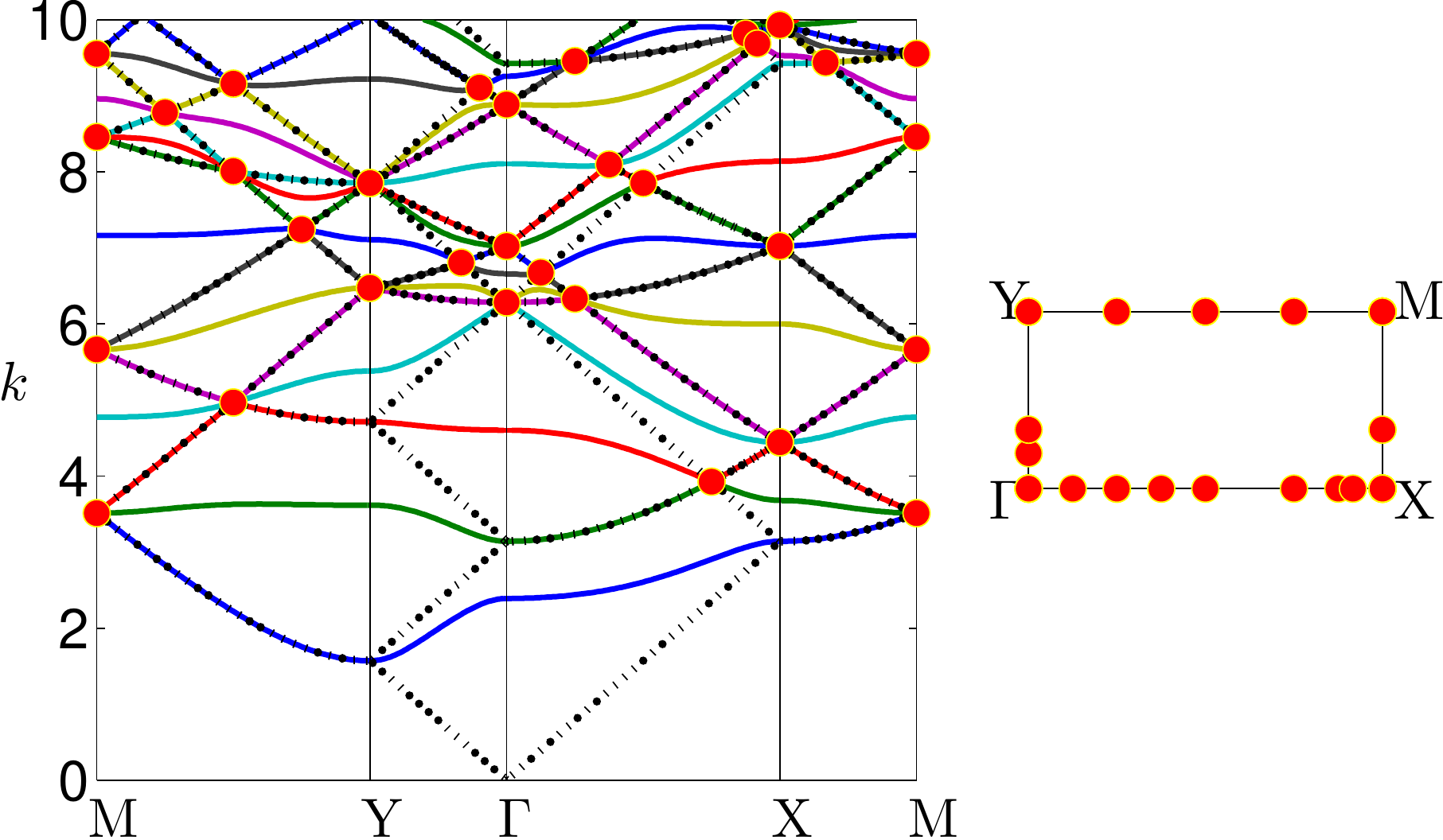}
\caption{\label{fig:bdiag4}}
\end{subfigure}
\caption{The band diagrams for \subref{fig:bdiag3} $\rho = \sqrt{3}$, and \subref{fig:bdiag4} $\rho = \sqrt{4}$, along the edges of the BZ for a rectangular array geometry.\label{bdiagb2}} 
\end{figure}

A selection of band diagrams for different aspect ratios is presented in Figures \ref{bdiagb1} to \ref{bdiagb2}. The aspect ratios chosen are square roots of integers, resulting in large numbers of 
triple intersections (not all at points of high symmetry in the Brillouin zone).

\section{Inertia of the elastic structure at Dirac points}

As commented in the Introduction, the physics of flexural waves differs from that of solutions to the Helmholtz or Schr\"{o}dinger equations. The latter are important in solid states physics and quantum mechanics, and in the literature, two types of structures associated with Dirac points are encountered. The first of these is said to correspond to massless particles, pertaining to solutions which correspond to an empty lattice. These band surfaces have the geometry of cones coming to a perfect point. The second type of solution discussed corresponds to rounded cones, with the associated particles being said to be massive, acquiring their mass through their interaction with the lattice.  For the solutions of the biharmonic equation considered in this paper, a different geometry occurs. The cone structures correspond to solutions sandwiched between light lines, and are associated with neutral lines, i.e., there is no interaction with the lattice. Once again, these may be described as massless. However, between the massless solutions converging to a perfect point, there exists a solution governed by the dispersion equation for flexural waves in the lattice \eqref{disprel2}. This is the massive solution. To distinguish this geometry for flexural waves, we will refer to the massless Dirac cones as being augmented by the massive solution. Note that the massive solution has a dispersion equation which guarantees that it intersects points of high symmetry in the Brillouin zone.

\begin{figure}[t!]
\centering
\begin{subfigure}[c]{.45\textwidth}
\includegraphics[width=\textwidth]{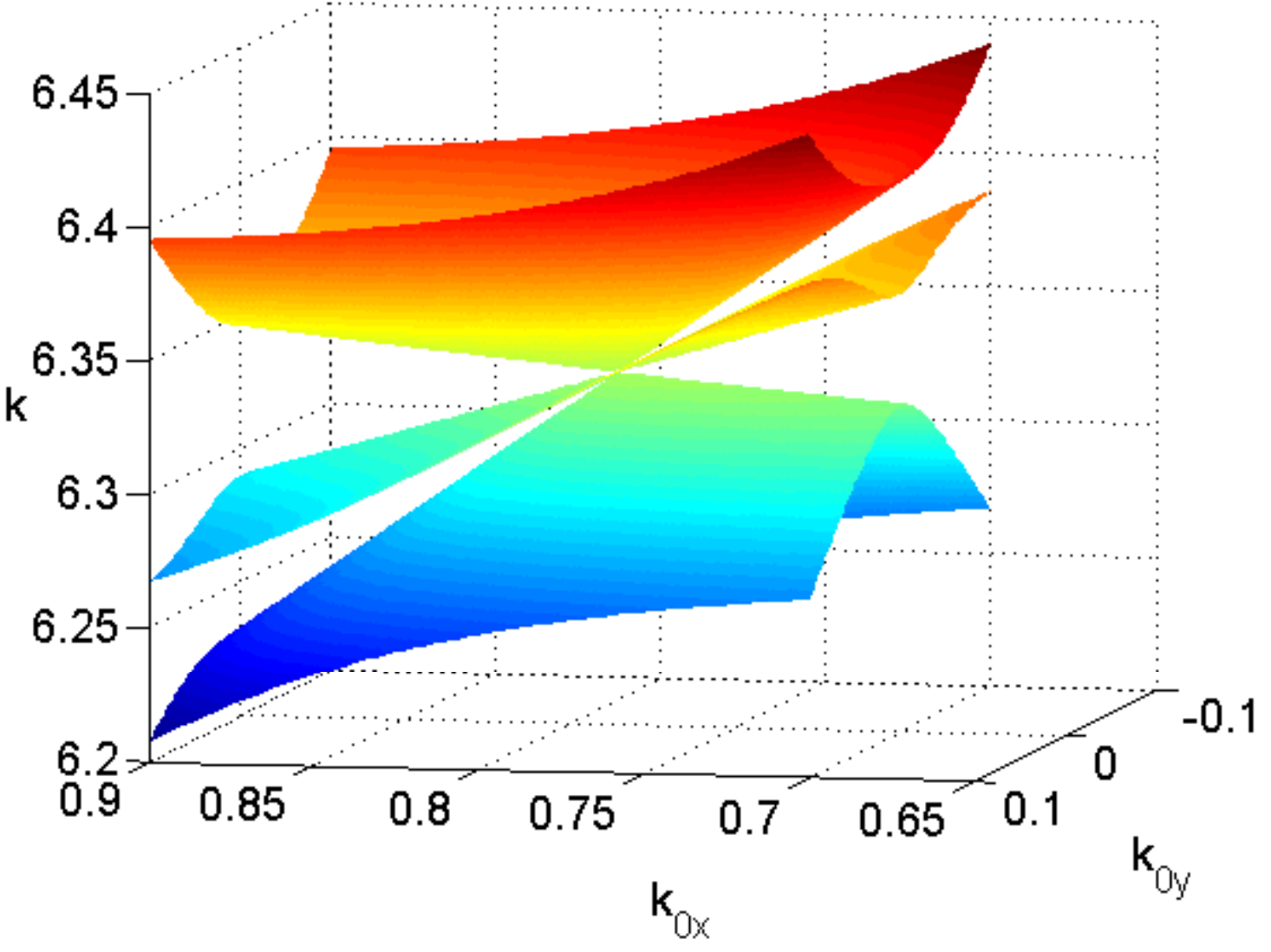}
\caption{\label{fig:rsdph}}
\end{subfigure} 
\begin{subfigure}[c]{.45\textwidth}
\includegraphics[width=\textwidth]{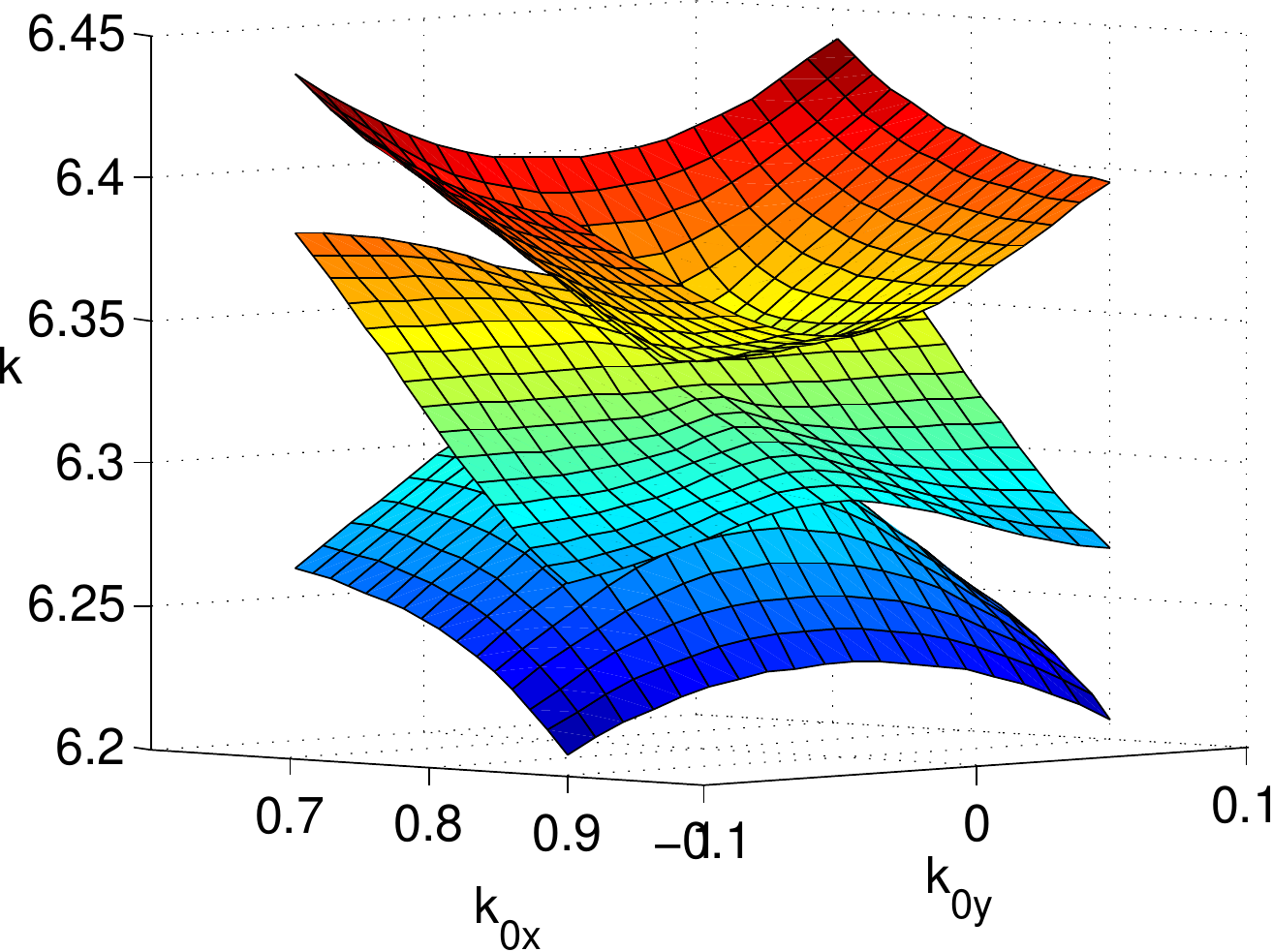}
\caption{\label{fig:rsdpm}}
\end{subfigure}
\caption{Two views of the dispersion surfaces near the  Dirac point at $k=6.3320$  located between the symmetry points $\Gamma$ and $\mathrm{X}$ at $(k_{0x},k_{0y}) = (0.7828,0)$, corresponding to an aspect ratio of $\rho = 2$. The central band is sloped at the Dirac point in contrast to Dirac points of high symmetry. \label{slantedDP}} 
\end{figure}

\begin{figure}[t!]
\centering
\begin{subfigure}[c]{.45\textwidth}
\includegraphics[width=\textwidth]{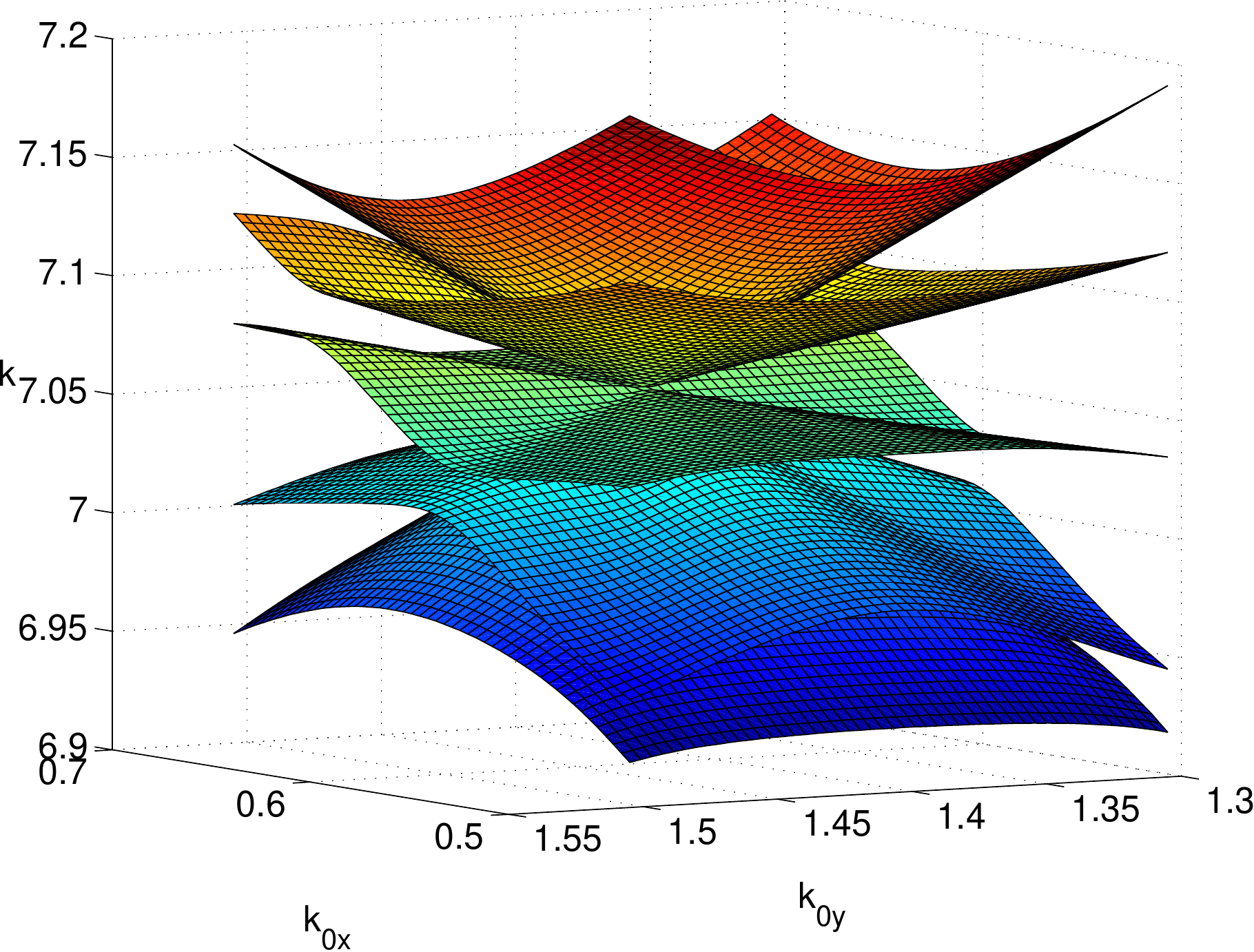}
\caption{\label{fig:bsurf1}}
\end{subfigure} 
\begin{subfigure}[c]{.45\textwidth}
\includegraphics[width=\textwidth]{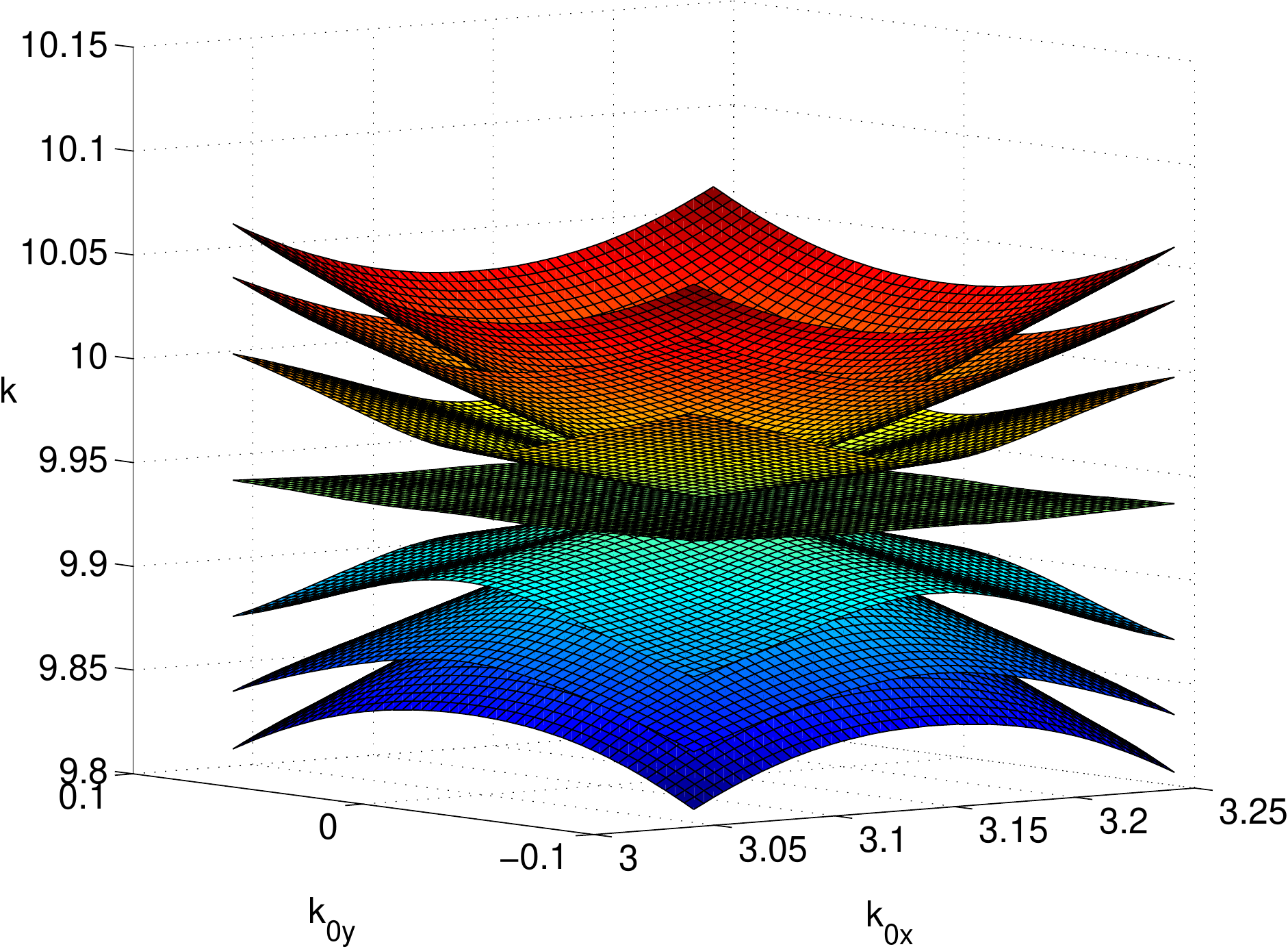}
\caption{\label{fig:bsurf2}}
\end{subfigure}
\caption{Three dimensional band surfaces in the vicinity of \subref{fig:bsurf1} $\mathbf{k}_0 = (0.6283,\pi/\sqrt{5})$ for $\rho = \sqrt{5}$, demonstrating a double Dirac-like cone away from a point of symmetry and \subref{fig:bsurf2} a triple Dirac-like cone at $\mathbf{k}_0 = \Xa$ for $\rho = 2$. \label{bsurfb1}} 
\end{figure}

When viewed on a three-dimensional diagram,  the surfaces corresponding to  massive solutions, such as those in figures  \ref{slantedDP} and  \ref{bsurfb1}, have at least one vanishing derivative of $\omega$  with respect to components of the Bloch vector at the points of intersection (this property is not readily visible on conventional dispersion diagrams confined purely to lines of high symmetry). In essence, the group velocities of the massive bands go  to zero, with the finite values for the density of states for Dirac cones given in table \ref{table1} (see equation \eqref{eq:dos5} for a discussion of the impact of this on the DOS).  In summary, the topology of what we term an augmented Dirac point for flexural waves has at least one massive central band and massless bands with frequencies above and below. This topology is present for flexural waves since they obey a fourth-order differential equation, rather than the second order Schr\"{o}dinger or Helmholtz equations. An interesting feature of the massless and massive bands is that the distinction between them is directionally dependent in the Brillouin zone. For example in Figure \ref{fig:bdiag2} corresponding to an aspect ratio of $\rho = \sqrt{2}$ at $k(Y) \approx  6.8$ we have, reading from bottom to top in $k$ for the segment $\mathrm{M} \mathrm{Y}$ the bands being: neutral, massive, neutral, massive and neutral. Reading in the same order along $\mathrm{Y} \Gamma$ the bands are: massive, neutral, massive, neutral and massive. The manner of construction for this typical band diagram makes this change of order appear to be sudden, whereas if viewed on a three-dimensional band surface, it is of course, gradual. In table 2 we summarise the key data describing the movement of Dirac cones along symmetry lines in the Brillouin zone for aspect ratios of the rectangular lattice varying from $\sqrt{1}$ to $\sqrt{10}$.
 
 \begin{table}[t]
	\centering
		\begin{tabular}{cccccc}
			 $\rho$ &  $\mathbf{k}_0$ & $k$ & $\simeq k$ & Trajectories & Cone type \\  \hline 
			$\sqrt{1}$ & $\Ma$ & $\sqrt{10}\pi$ & $9.9346$ & one along $\Ma \Xa$ & double \\ \hline
			$\sqrt{2}$ & $\Ya$ & $3\pi/\sqrt{2}$ & $6.6643$ & one along $\Ya \Ma$ & double \\ \hline
			$\sqrt{2}$ & $\Xa$ & $3\pi$ & $9.4248$ & one along $\Xa \Ma$ & double \\ \hline 
			$\sqrt{3}$ & $\Ga$ & $4\pi/\sqrt{3}$ &  $7.2552$ & one along $\Ga \Xa$ & double \\ \hline
			$\sqrt{3}$ & $\Ma$ & $2\pi/\sqrt{7/3}$ & $9.5977$ & two along $\Ma \Xa$ and two along $\Ma \Ya$ & triple \\ \hline
			$\sqrt{4}$ & $\Xa$ & $ \sqrt{10} \pi$ &  $9.9346$ & two along $\Xa \Ma$ and two along $\Xa \Ga$ & triple \\ \hline
		  $\sqrt{4}$ & $\Ya$ & $ 5\pi / 2$ &  $7.8540$ & one along $\Ya \Ma$  & double \\ \hline
			$\sqrt{5}$ & $\Ga$ & $6\pi/\sqrt{5}$ & $8.4298$ & one along $\Ga \Xa$ & double \\ \hline
			$\sqrt{5}$ & \multicolumn{5}{c}{{\it Special case: double at $k=7.0529$ between $\Ma \Ya$ at $\mathbf{k}_0 = (0.6283,\pi/ \sqrt{5} )$}} \\ \hline
			$\sqrt{6}$ & $\Ma$ & $ \sqrt{55/6}\pi$ & $9.5116$ & two along $\Ma \Ya$ and two along $\Ma \Xa$ & triple \\ \hline
			$\sqrt{6}$ & $\Ya$ & $ 5\pi /\sqrt{6}$ & $6.4127$ & one along $\Ya \Ma$ & double \\ \hline
			$\sqrt{6}$ & $\Ya$ & $ 7\pi /\sqrt{6}$ & $8.9778$ & one along $\Ya \Ma$ & double \\ \hline
			$\sqrt{7}$ & $\Ga$ & $ 8\pi /\sqrt{7}$ & $9.4993$ & one along $\Ga \Xa$ & double \\ \hline
			$\sqrt{8}$ & \multicolumn{5}{c}{{\it Special case: double at $k=7.8146$ between $\Ma \Ya$ at $\mathbf{k}_0 = (0.7853, \pi/\sqrt{8})$}} \\ \hline
			$\sqrt{8}$ & $\Ga$ & $ 3\pi /\sqrt{2}$ & $6.6643$ & one along $\Ga \Xa$ & double \\ \hline
			$\sqrt{8}$ & $\Ya$ & $ 9\pi / (2 \sqrt{2} ) $ & $9.9965$ & one along $\Ya \Ma$ & double \\ \hline
			$\sqrt{8}$ & $\Xa$ & $ 3\pi$ & $9.4248$ & one along $\Xa \Ga$ & double \\ \hline
			$\sqrt{9}$ & $\Ma$ & $ \sqrt{10}\pi$ & $9.9346$ & two along $\Ma \Xa$ and two along $\Ma \Ya$ & triple \\ \hline
			$\sqrt{10}$ & $\Ma$ & $ \sqrt{91/10}\pi$ & $9.4770$ & two along $\Ma \Ya$ and two along $\Ma \Xa$ & triple \\ \hline
		\end{tabular}
	\caption{A summary of aspect ratios corresponding to the start of Dirac cone transits along the edge of the Brillouin Zone for $k<10$. Note that `double' refers to a double Dirac cone at the starting point which gives rise to a single Dirac cone transit, and similarly `triple'  (Triple Dirac cones) coincides with four Dirac cone transits. Futhermore, $AB$ denotes a transit from the symmetry point $A$ to the symmetry point $B$.}
	\label{tab:TDCandDDC}
\end{table}

\section{Unidirectionally localised vibration modes in platonic clusters}

According to Craster {\it et al.} \cite{craster2010high}, the notion of high-frequency homogenisation is introduced with  reference to a critical point on the dispersion surface where the group velocity is equal to zero. Given a standing wave frequency, both the Bloch vector and the frequency are perturbed and an asymptotic solution is written in the form of a product of the rapidly oscillating standing wave and a smooth envelope function, which satisfies a partial differential equation,  referred to as an homogenised equation. It was also noted in \cite{movchan2014resonant}  that this equation does not have to be elliptic. In other words, the homogenised solution may exhibit  localisation along some preferred directions. In particular, in  \cite{ayzenberg2008resonant}  such solutions are referred to as star-shaped wave forms, which corresponds to a hyperbolic homogenised equation for the envelope function. Furthermore, the preferred directions are aligned with the characteristics of this hyperbolic equation.

\begin{figure}[t!]
\centering
\begin{subfigure}[c]{0.45\textwidth}
\includegraphics[width=\linewidth]{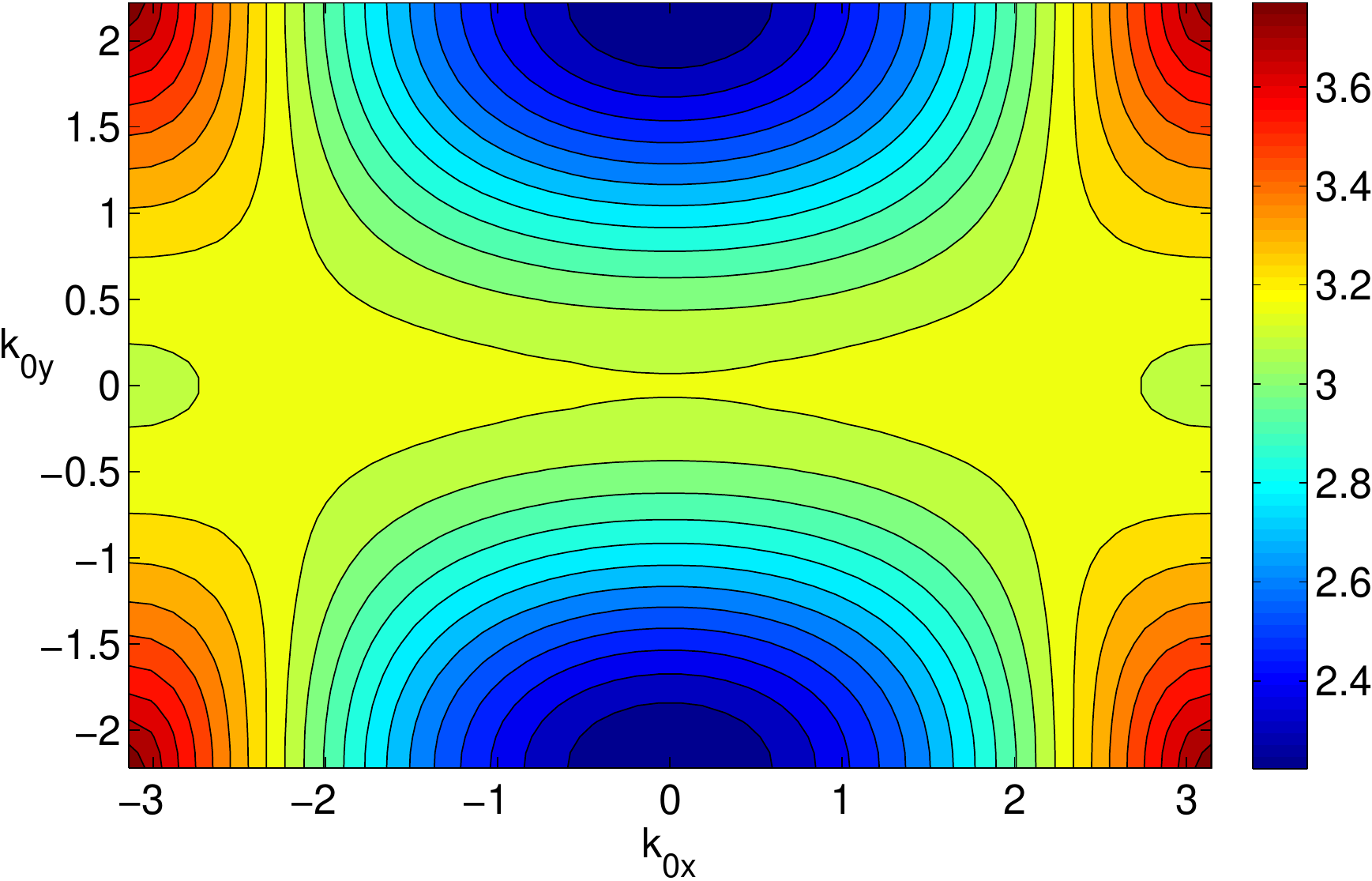}
\caption{\label{fig:fig14a}}
\end{subfigure}
\qquad
\begin{subfigure}[c]{0.45\textwidth}
\includegraphics[width=\linewidth]{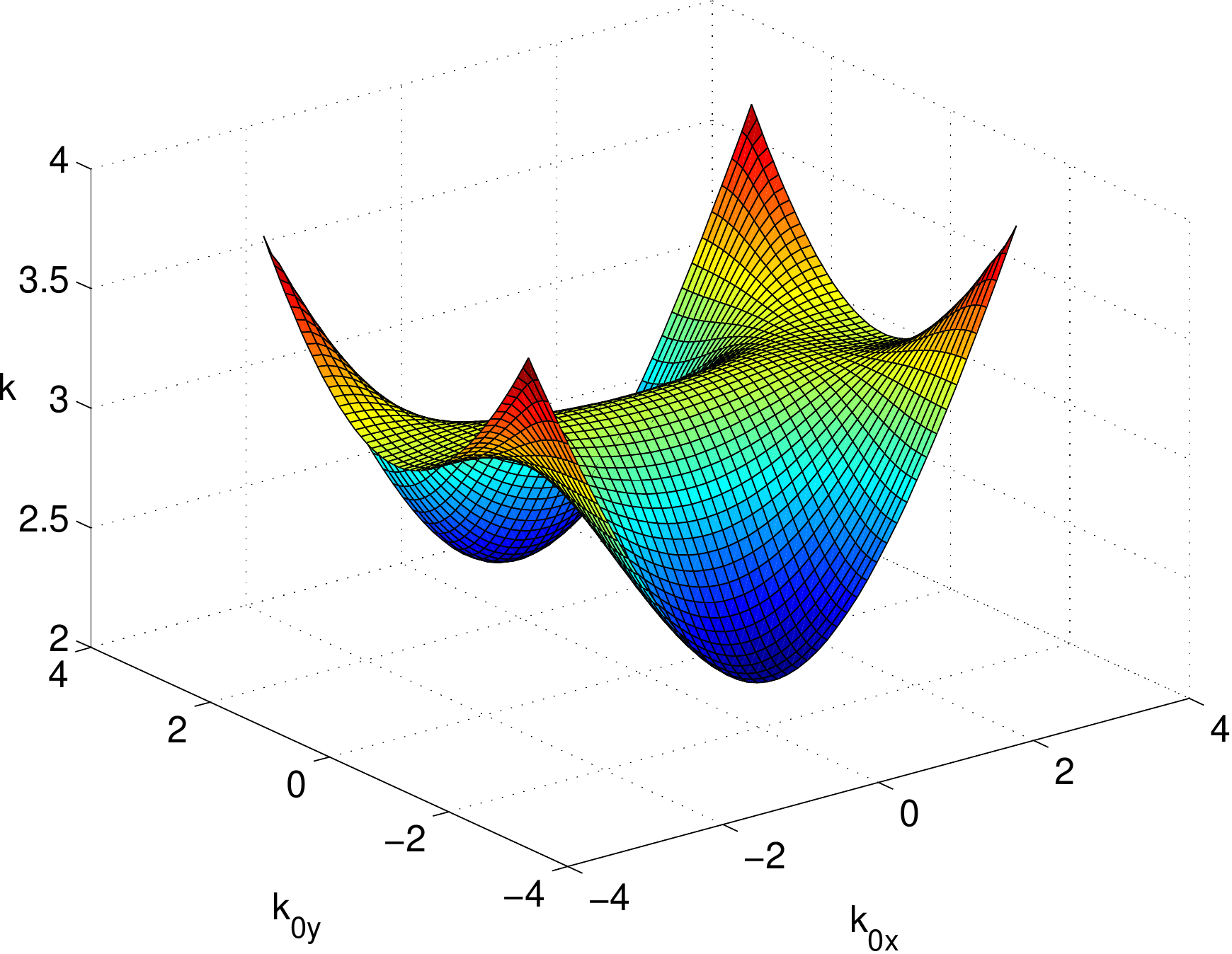}
\caption{\label{fig:fig14b}}
\end{subfigure} \\
\begin{subfigure}[c]{0.45\textwidth}
\includegraphics[width=\linewidth]{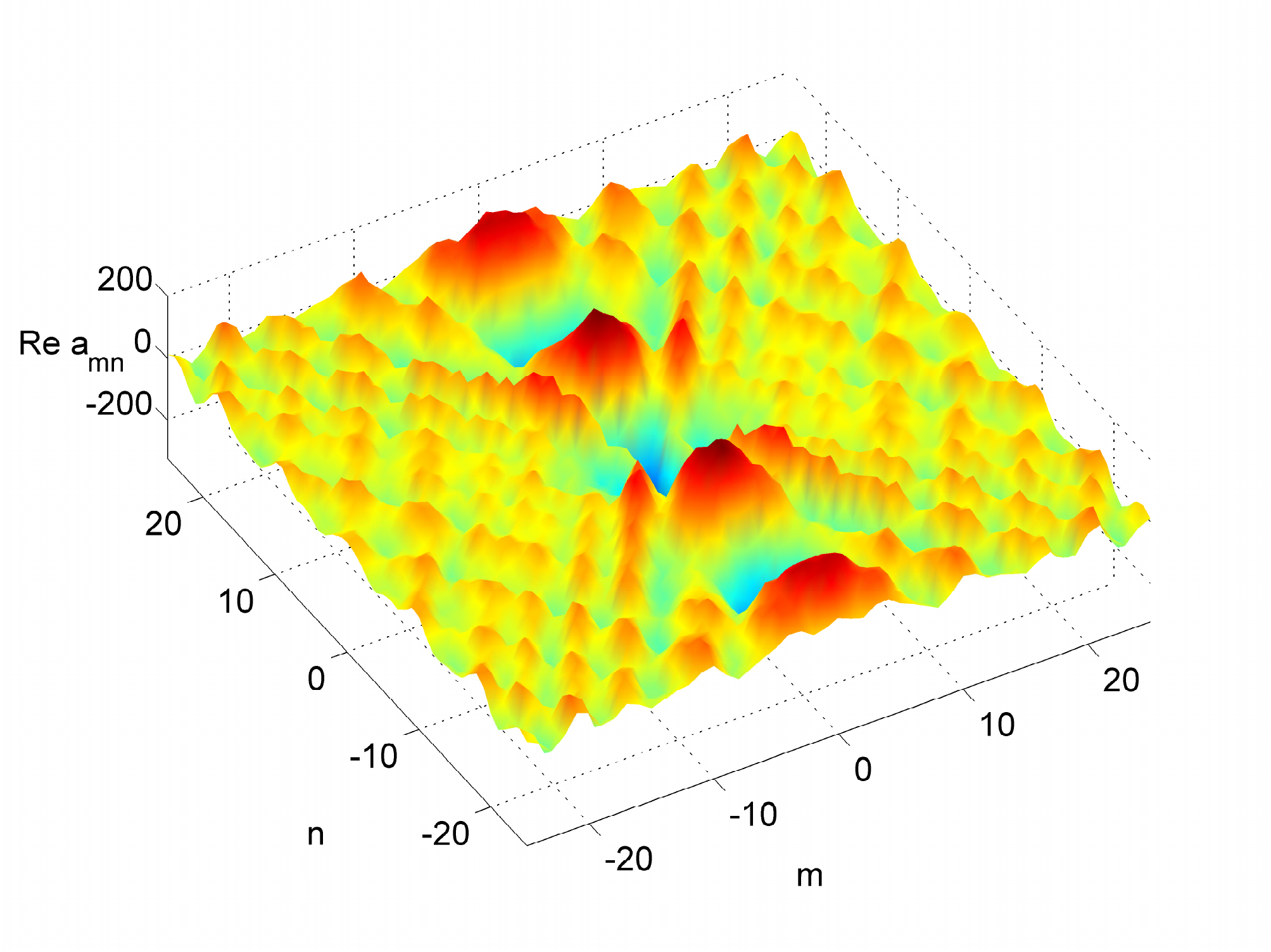}
\caption{\label{fig:amnsqrt2parab}}
\end{subfigure}
\caption{The \subref{fig:fig14a} contour and \subref{fig:fig14b} surface plot for the first dispersion surface corresponding to $\rho = \sqrt{2}$ over the entire Brillouin zone. The $a_{mn}$ coefficients showing the forces imposed (to enforce pinned conditions) at $k=3.069$ are given in  \subref{fig:amnsqrt2parab} which corresponds to a localised mode along the $y$ axis.\label{fig13}} 
\end{figure}

In the framework of the approach presented here, we have discovered a new phenomenon for flexural elastic waves, corresponding to envelope functions that satisfy homogenised equations of the parabolic type. This effect has never been addressed in the past for flexural plate structures (however such behaviour, referred to as the `line-defect' effect, has been observed previously in metamaterials  \cite{antonakakis2013asymptotics}). There is also an apparent connection with the position of Dirac cones. Namely, a parabolic cylindrical profile of the dispersion surface occurs in close proximity to Dirac cone points. The corresponding wave forms can propagate freely along the direction of the generator of the cylinder, and are localised in the orthogonal direction. To illustrate this finding, we refer to the dispersion diagram in figure \ref{fig:bdiag2}, where Dirac cone points are clearly identified, as well as stationary points where the group velocity of Bloch waves in an infinite periodic structure is equal to zero. First, an accurate study of the full three-dimensional dispersion surface in the first pass band reveals a parabolic profile at the frequency of $k = 3.1538$. This surface is displayed in figure   \ref{fig13}, and it is also accompanied by the contour line diagram on the inset of the same figure. 
An independent test has been performed for a sufficiently large but finite cluster, which has been `cut out' of the doubly periodic system. A frequency response problem has been studied for the case when a finite time-harmonic displacement has been applied to the central pin of the cluster whereas the remaining pins of the cluster remain fixed. 
 The solution method for finite pinned platonic crystals, or platonic clusters \cite{mcphedran2009platonic,meylan2011fast,smith2012negative} is as follows. For the problem of a plate which is constrained at a set of points, where the    pin  at $\mathbf{x}_{00}$  is forced, we can express the displacement in the form
\begin{equation}
w(x,y) = g(\mathbf{x} ,\mathbf{x}_{00}) +   
\sum_{\substack{m,n=-M\\(m,n) \neq (0,0)}}^{M} 
 a_{mn} g(\mathbf{x} , \mathbf{x}_{mn}),  
\label{eq:finitew}
\end{equation}
where   $\mathbf{x}_{mn} =(m d_x, n d_y)$. The free-space Green's function for the biharmonic operator is given by
\begin{equation}
g(\mathbf{x},\mathbf{x}^\prime) = \frac{\rmi}{8k^2} \left[ H_0^{(1)}(k|\mathbf{x} - \mathbf{x}^\prime|) - \frac{2}{\pi \rmi} K_0(k|\mathbf{x} - \mathbf{x}^\prime|) \right],
\end{equation}
which satisfies 
\begin{equation}
\left(\Delta^2 - k^4 \right) g(r) = \delta(r).
\end{equation}
 \begin{figure}[t!]
\centering
\begin{subfigure}[c]{0.45\textwidth}
\includegraphics[width=\linewidth]{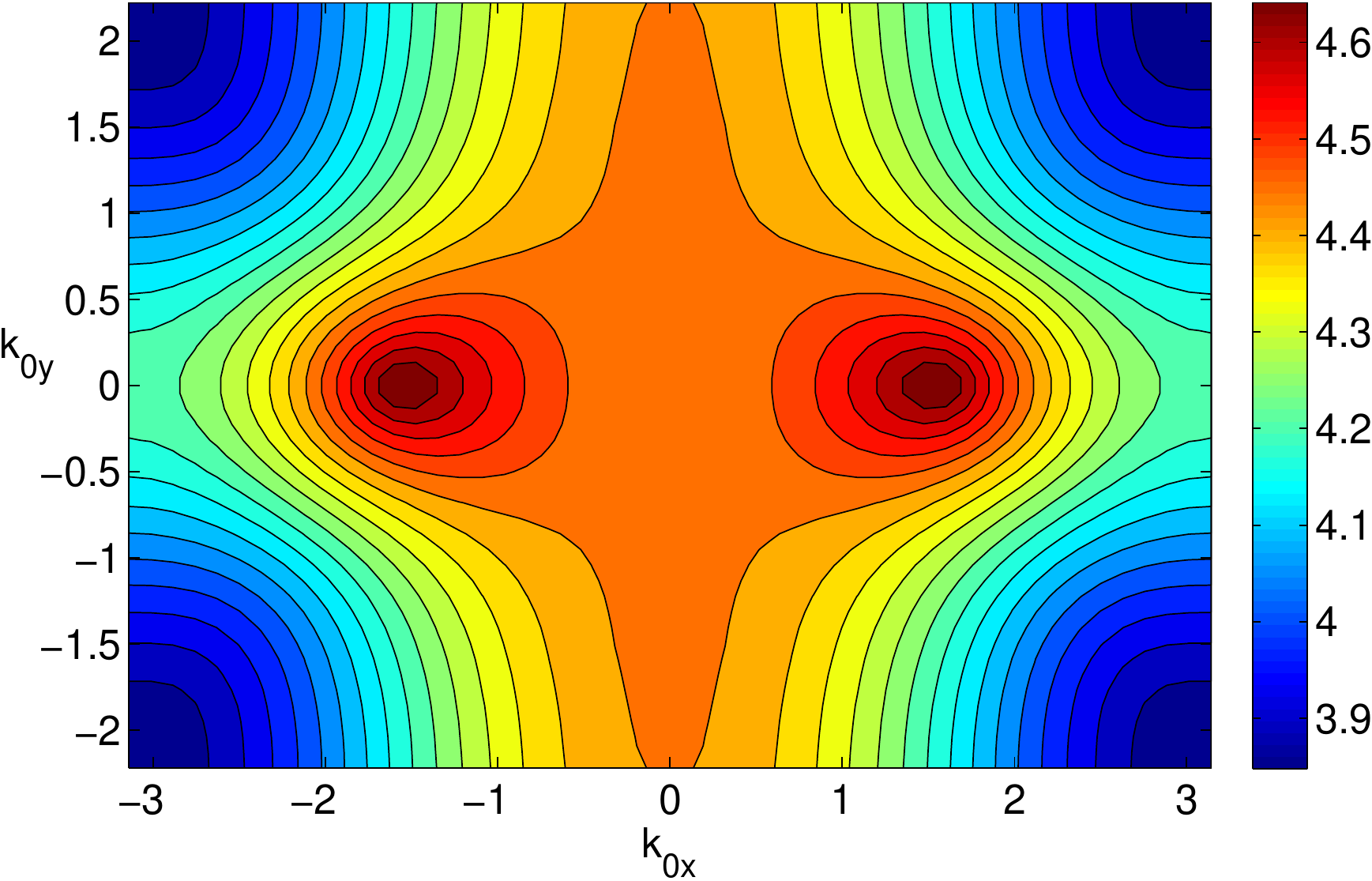}
\caption{\label{fig:fig15a}}
\end{subfigure}
\qquad
\begin{subfigure}[c]{0.45\textwidth}
\includegraphics[width=\linewidth]{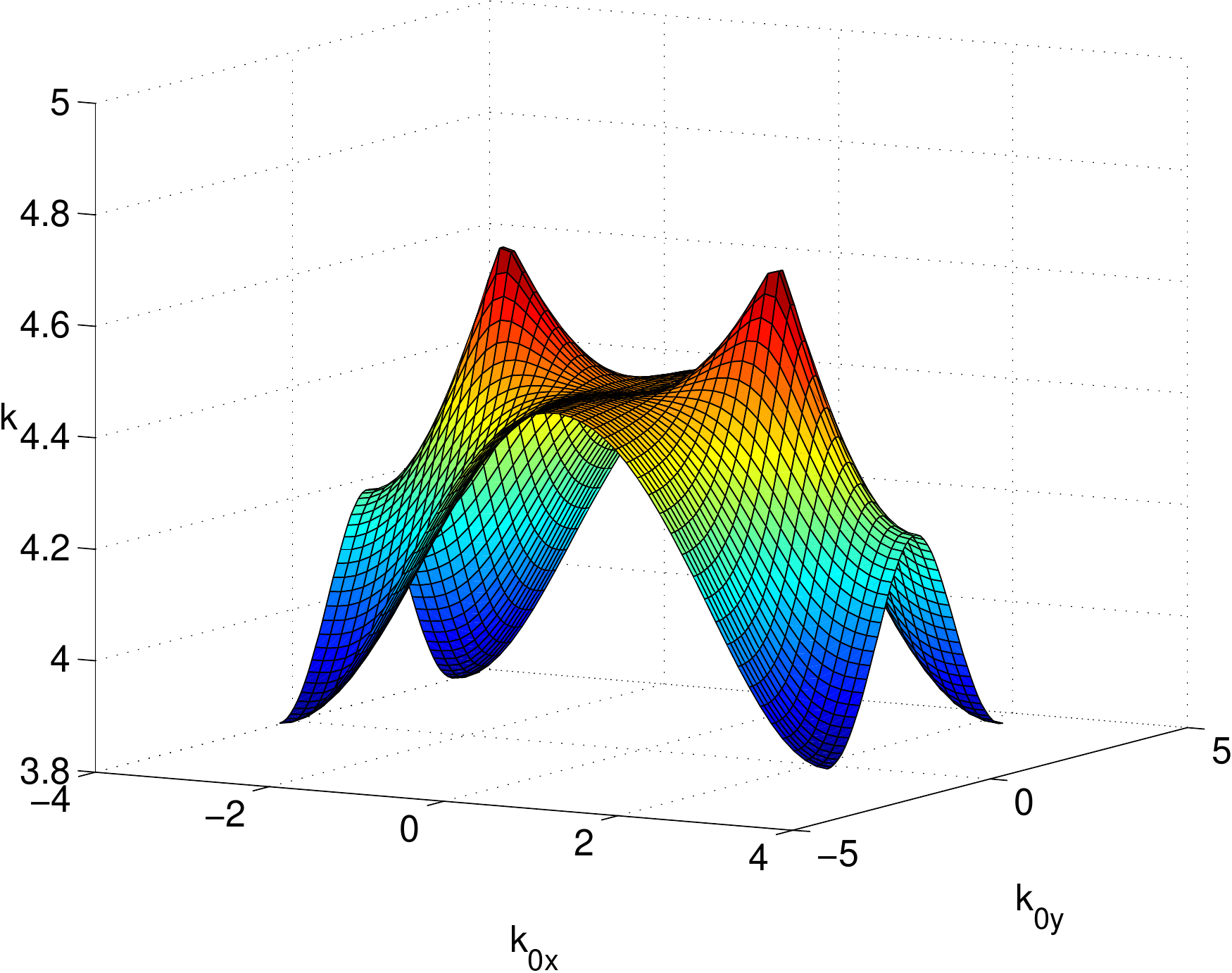}
\caption{\label{fig:fig15b}}
\end{subfigure} \\
\begin{subfigure}[c]{0.45\textwidth}
\includegraphics[width=\linewidth]{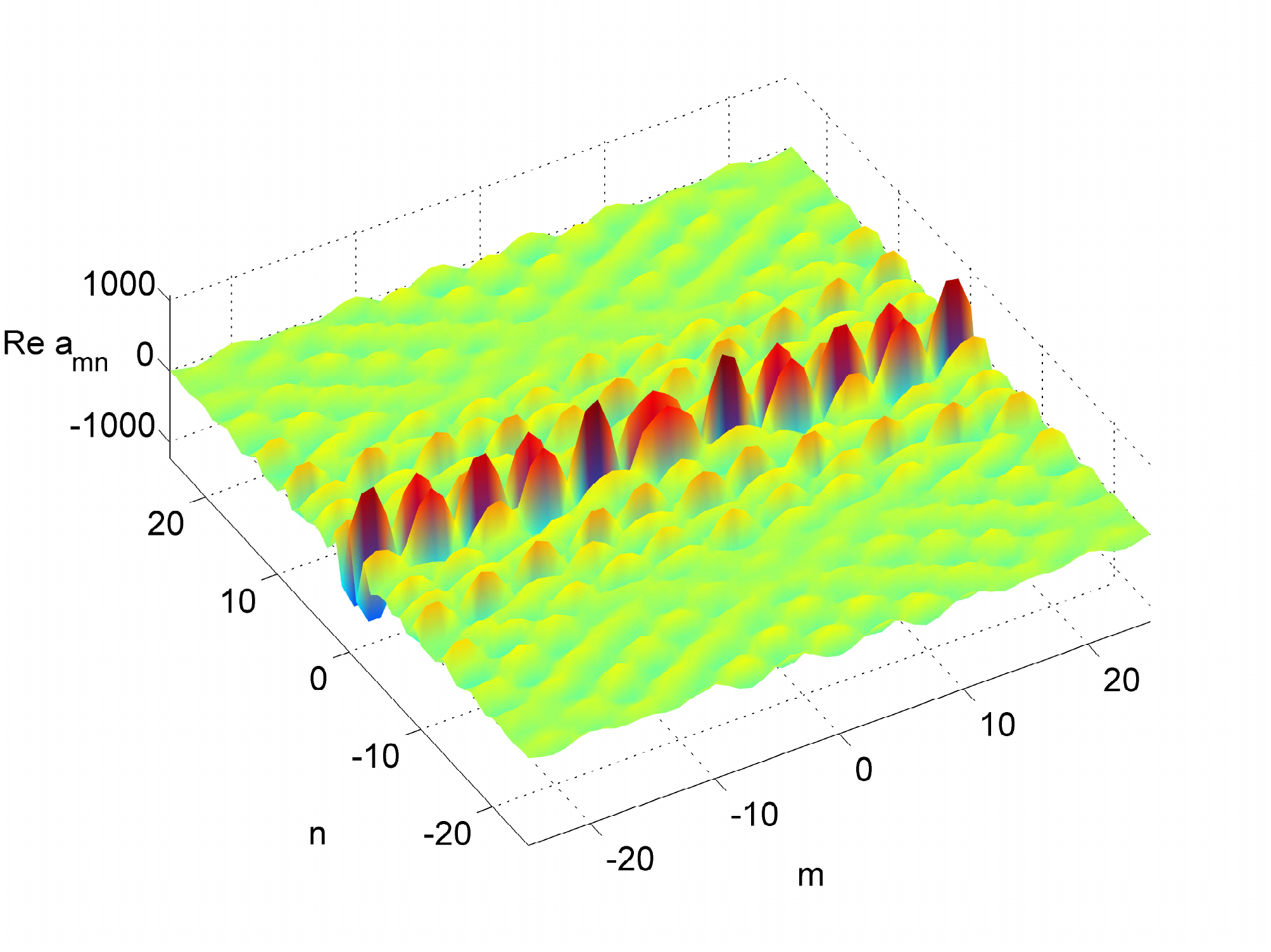}
\caption{\label{fig:amnsqrt2parab2}}
\end{subfigure}
\caption{The \subref{fig:fig15a} contour and \subref{fig:fig15b} surface plot for the second dispersion surface corresponding to $\rho = \sqrt{2}$ over the entire Brillouin zone. The $a_{mn}$ coefficients showing the forces imposed (to enforce pinned conditions) at $k=4.4000$ are given in  \subref{fig:amnsqrt2parab2} which corresponds to a localised mode along the $x$ axis.\label{fig15}} 
\end{figure}
We then specify the boundary conditions $w(\mathbf{x}_{pq}) = \delta_{p0}\delta_{q0}$ to obtain the linear system
\begin{equation}
\mathbf{A} \mathbf{a} = \mathbf{f},
\label{eqfinitesystem}
\end{equation}
where $A_{ij} = g(\mathbf{x}_{i},\mathbf{x}_{j})$ which refers to the index pairs $(m,n)$ and $(p,q)$ and $f_j = \delta_{p0}\delta_{q0}$, for the unknown $a_{mn}$ coefficients. Using \eqref{eq:finitew}  we are able to generate  displacement fields for different finite clusters which possess an interior rectangular lattice geometry. For a suitably chosen geometry and size we are able to demonstrate the features of an infinite array of pinned points. Thus, the forces $a_{mn}$ have been computed and then plotted at every pin of the structure, as shown in figure \ref{fig:amnsqrt2parab}. For the case of a uni-directional localisation, the spikes of the force values are concentrated along a single line. The orientation of  localised modes  depends on the position of the neighbouring Dirac points in reciprocal space. The form of the localised modes does not depend on the choice of the source point positioned properly within that cluster. The above statement becomes evident when we refer to figure \ref{fig15}, which contains a dispersion surface and a uni-directional localised mode at the frequency of $k=4.4$. The locally parabolic surface occurs inbetween two conical regions, and the corresponding frequency of the uni-directional mode is just below the frequency value identified for the vertex of the Dirac cone. The corresponding localised mode is also shown on the inset of the same figure. It is worth mentioning that the orientations of the uni-directional modes in figures \ref{fig13} and \ref{fig15} are mutually orthogonal to each other. It will be interesting to investigate the correspondence between uni-directional modes in finite clusters and  surface waves on semi-infinite flexural structures.

\section*{Funding statement}
{\small R.C.M., A.B.M., N.V.M. and M.B.  gratefully acknowledge financial support from the European Community's Seven Framework Programme under contract numbers PIAP-GA-2011-286110-INTERCER2, 
PIAPP-GA-28544-PARM-2
and PIEF-GA-2011-302357-DYNAMETA. 
M.J.A.S. would like to acknowledge the support of Dr Sebastien Guenneau and resources provided through his ERC starting grant ANAMORPHISM.}

\section*{Appendix A: Description of the animations}
\label{app:1}
 
\centering
 
\begin{enumerate}[i]
\item Animation 1 (\href{http://dx.doi.org/10.6084/m9.figshare.1186824}{doi:10.6084/m9.figshare.1186824}): {\it In this animation we observe the evolution of the band diagram as we alter the aspect ratio $\rho$ of our rectangular array of pinned points through the range $1<\rho< \sqrt{10}$. Also presented here are the   the light line surfaces along the edges of the Brillouin zone. It is clearly apparent that there are Dirac point transits along the edges of the zone, which come together when $\rho$ takes square root integer values. One other interesting feature observed include the closing of the low frequency band gap with increasing aspect ratio.}

\item Animation 2 (\href{http://dx.doi.org/10.6084/m9.figshare.1186825}{doi:10.6084/m9.figshare.1186825}): {\it In this animation we present the same band diagram evolution as in Animation 1, except over the shorter interval $1<\rho<3$. Here we superpose the triple point intersections given in \eqref{degll8} and \eqref{degll9} and accompany the diagram with the top-down projection of these points along the edges of the Brillouin zone. This animation follows the style of figures \ref{bdiagb1}  and \ref{bdiagb2}, and a summary of the evolution of these Dirac points is given in table  \ref{tab:TDCandDDC}.}

\item Animation 3 (\href{http://dx.doi.org/10.6084/m9.figshare.1186826}{doi:10.6084/m9.figshare.1186826}) {\it  In this animation we present the first few three-dimensional band surfaces for a rectangular array of pinned points corresponding to the aspect ratio $\rho = \sqrt{2}$. This movie presents the dynamic and ever-changing behaviour of the slowness surfaces (isofrequency contours for $\omega = k^2/\sqrt{\rho_m h/D}$) as we move through increasing wave number $k$ (as observed through the cut plane advancing through $k$ in the right-hand figure).

}

\item Animation 4 (\href{http://dx.doi.org/10.6084/m9.figshare.1186827}{10.6084/m9.figshare.1186827}) {\it This animation tracks the tilting of a single Dirac cone (with central flat band) as it transits between symmetry points for  increasing aspect ratio values between $1\leq \rho\leq\sqrt{2}$. This cone starts at a double Dirac cone at $\mathrm{M}$ and terminates  at  at $\mathrm{X}$ to form a double Dirac point, of which we see the central and two upper bands in the final moments of the simulation (see table  \ref{tab:TDCandDDC} for further details).

}

\end{enumerate}

\end{document}